\documentclass{jfm}

\usepackage{graphicx}
\usepackage{newtxtext}
\usepackage{newtxmath}
\usepackage{natbib}
\usepackage{hyperref}
\usepackage{multirow}
\hypersetup{
    colorlinks = true,
    urlcolor   = blue,
    citecolor  = black,
}

\newcommand{\RomanNumeralCaps}[1]
\linenumbers

\newcommand{\ii}{\mathrm{i}}
\newcommand{\ee}{\mathrm{e}}

\title{On the preferred flapping motion of round twin jets}

\author{Daniel Rodr\'iguez\aff{1} \corresp{\email{daniel.rodriguez@upm.es}},
Michael N. Stavropoulos\aff{2},
Petr\^onio A. S. Nogueira\aff{2},
Daniel M. Edgington-Mitchell\aff{2}
\and Peter Jordan\aff{3} }

\affiliation{\aff{1}Universidad Polit\'ecnica de Madrid, ETSIAE-UPM, Plaza del Cardenal Cisneros 3, 28040 Madrid, Spain.
\aff{2}Department of Mechanical and Aerospace Engineering, Laboratory for Turbulence Research in
Aerospace and Combustion, Monash University, Clayton 3800, Australia.
\aff{3}D\'epartement Fluides Thermique et Combustion, Institut Pprime-CNRS-Universit\'e de Poitiers-ENSMA, 86962 Chasseneuil-du-Poitou, Poitiers, France.}

\begin{document}
\maketitle

\begin{abstract}
Linear stability theory (LST) is often used to model the large-scale flow structures in the turbulent mixing region and near pressure field of high-speed jets. For perfectly-expanded single round jets, these models predict the dominance of $m=0$ and $m = 1$ helical modes for the lower frequency range, in agreement with empirical data. 
When LST is applied to twin-jet systems, four solution families appear following the odd/even behaviour of the pressure field about the symmetry planes. The interaction between the unsteady pressure fields of the two jets also results in their coupling. 
The individual modes of the different solution families no longer correspond to helical motions, but to flapping oscillations of the jet plumes. In the limit of large jet separations, when the jet coupling vanishes, the eigenvalues corresponding to the $m=1$ mode in each family are identical, and a linear combination of them recovers the helical motion. Conversely, as the jet separation decreases, the eigenvalues for the $m=1$ modes of each family diverge, thus favouring a particular flapping oscillation over the others and preventing the appearance of helical motions.
The dominant mode of oscillation for a given jet Mach number $M_j$ and temperature ratio $T_R$ depends on the Strouhal number $St$ and jet separation $s$. Increasing both $M_j$ and $T_R$ independently is found to augment the jet coupling and modify the $(St,s)$ map of the preferred oscillation mode. Present results predict the preference of two modes when the jet interaction is relevant, namely varicose and especially sinuous flapping oscillations on the nozzles plane.
\end{abstract}



\section{Introduction}
\label{sec:Introduction}

Tactical fighters developed since the 1960s predominately feature fuselage-embedded twin jet engines. Additionally, multi-tube nozzles have been investigated as possible jet noise suppressors, and designs of future distributed-propulsion systems involve placing two or more parallel jet streams in close proximity. The closely spaced jets can interact both at the hydrodynamic and acoustic levels, giving rise to complex flow structures when compared with single round jets at equivalent operating conditions. 

Following the seminal works of \citet{Mollo:JAM67}, the relation between the dominant components of the far-field noise radiated by high-speed jets and large-scale fluctuations in the turbulent mixing region, coherent over several nozzle diameters, has been the subject of growing research \citep{JordanColonius:ARFM13,Cavalieri:AMR2019}.  Radiated sound is correlated with large-scale, low-frequency fluctuations in the mixing region and with very few azimuthal modes for single isolated jets \citep{Juve:JSV80,Hileman:JFM05,Cavalieri:JSV11b}. The existence of coherent structures in turbulent jets was first identified by \citet{CrowChampagne:JFM71} and their resemblance to instability waves for harmonically-forced supersonic jets suggested the use of linear instability analysis to model them \citep{CrightonGaster:JFM76,MichalkePrAS1984}. The presence of wavepackets in high-speed jets was finally demonstrated over the last two decades, as well as the ability of linear stability calculations to model them faithfully \citep{Suzuki:JFM06,Gudmundsson:JFM11,Cavalieri:JFM13}. 

As opposed to single round jets, instability analyses for twin jets are scarce in the literature, due to the mathematical complexity of the latter. The mean turbulent flow corresponding to isolated round jets is axially symmetric, enabling the introduction of azimuthal Fourier modes (each one characterised by an integer wavenumber $m$) and only requires spatial discretisation in the radial direction. Bipolar coordinates were used by \citet{Morris:JFM90} to study the inviscid instability of two axially-homogeneous parallel jets. He identified the counterparts of the different azimuthal Fourier modes known for single jets and classified them according to the symmetries about the jet-center plane and the plane normal to it. \citet{GreenCrighton:JFM97} used a similar approach to study coupled oscillation modes of the jet cores, considering varicose and sinuous flapping motions of the two jets. More recently, \citet{Rodriguez:CRM2018} and \citet{Nogueira:JFM21} analysed the local linear instability of twin-jet configurations by applying two-dimensional cross-stream discretisations that do not restrict the spatial structure of the wavepackets. Interestingly, the latter analyses recover the same families of eigenmodes corresponding to the Fourier modes of single jets, albeit modified on account of the jet-jet interaction; additional eigenmodes corresponding to mechanisms not present for single jets were not identified.
It was observed (but not reported) that the eigenfunctions for $m > 0$ modes do not correspond to helical oscillations, but constitute combinations of the respective $+m$ and $-m$ helical modes with the precise phase relationship such that they describe flapping motions on the lateral or vertical planes.

This is consistent with experimental observations and simulations in supersonic twin round jets. \citet{Seiner:AIAAJ88} and \citet{Wleizen:AIAAJ89} showed that the B-mode of oscillation associated with screech manifests as a coupled flapping motion of the plumes occurring in the plane containing the jets. \citet{Alkislar:AIAAJ2005} reported that the coupling between the two jets, for their particular conditions of jet Mach number $M_j$ and jet separation $s$, results in a symmetric flapping with respect to the mid plane and dominant jet oscillations occuring in the plane containing the jets (referred to as lateral flapping).
A similar observation was made by \citet{Kuo:AIAAJ17} based on Large Eddy Simulations.
 The experiments by \citet{Kuo:AIAAJ17} considered a twin-jet configuration with separation $s/D = 2$ and convergent-divergent nozzles with design Mach number of 1.23 and jet Mach numbers between $M_j = 1.15$ and 1.4. The jets exhibited coupled oscillations, with a varicose (symmetric) lateral flapping motion being observed at conditions in which a single jet would present the helical oscillations typical of the screech B-modes.
Experiments undertaken by \citet{Knast:AIAAJ18,Bell:EF18} reported both helical and flapping oscillations in twin jets, but due to experimental constraints the presence or absence of helical modes could not be rigorously determined. Later, \citet{Bell:JFM21} demonstrated that, even at a fixed operation condition, twin-jet oscillations present intermittency. At some time lapses the motion of the two jets can be uncorrelated and present helical motions; at other lapses they are strongly correlated and present a coupled lateral flapping.
 
This paper revisits the locally-parallel linear instability of twin-jet configurations. The first objective is to show that the coupling of the two jets favours flapping motions over helical ones. The analysis is focused on $m=1$ modes, as experiments show their prevalence over other modes for most flow conditions in the supersonic regime, specially in the first diameters from the nozzle lip. The second objective is to map the preferred flapping mode for each Strouhal number and jet separation, and the impact of the jet Mach number and temperature ratio upon them.
Two independent formulations of the linearised equations are used: (i) an approach that discretises the cross-stream plane using Cartesian coordinates, valid for of finite-thickness jets of arbitrary shape; and (ii) an inviscid vortex-sheet method analogous to that used by \citet{Morris:JFM90,DU1993,Stavropoulos:AIAA2021}. Section \ref{sec:Methodology} describes the two formulations. Section \ref{sec:Results} presents the results of the analyses. The relevant eigenmode families are described briefly. The impact of the interaction of the pressure fields of the two jets on the eigenmodes and the appearance of preferred modes is then discussed. A parametric study is then presented that analyses the effect of the jet Mach number and temperature ratio on the preferred oscillation mode. The main conclusions are summarised in section \ref{sec:Conclusions}.


\begin{figure}
\centerline{
\includegraphics[trim = 8cm 4cm 6.5cm 3cm, clip = true, width = .5\textwidth]{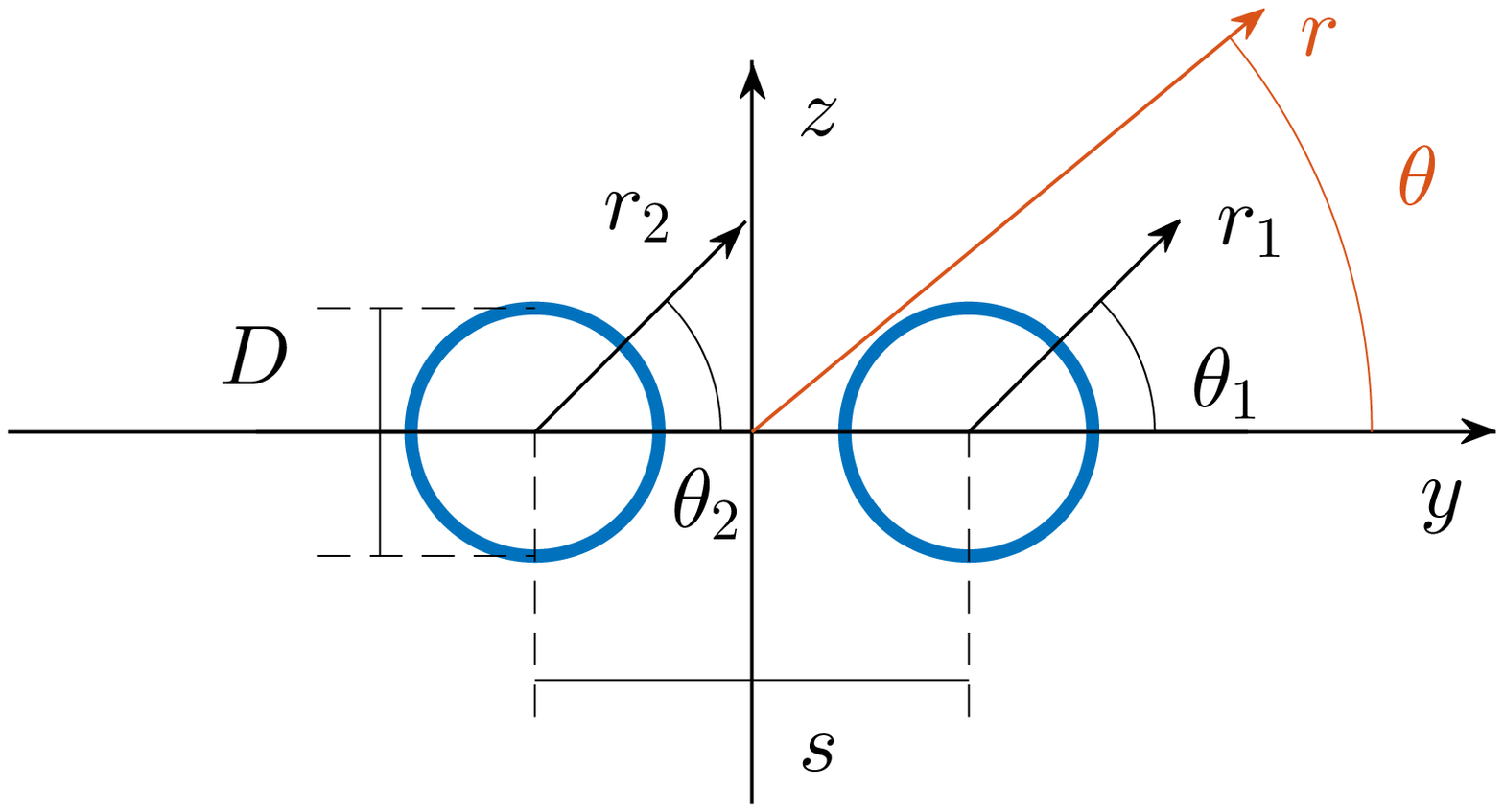}
}
\caption{Twin-jet configuration and geometry, showing the different coordinate systems employed.}
\label{fig:sketch}
\end{figure}

\section{Formulations of the linear stability problem for twin jets}
\label{sec:Methodology}
\vspace{5pt}

The twin-jet geometry and geometrical parameters are shown in figure \ref{fig:sketch}. The nozzle diameter is $D$ and the separation between the jet axes is $s$. The streamwise coordinate $x$ is oriented perpendicular to the paper towards the reader. Radial and azimuthal coordinates measured from the origin are denoted by $(r,\theta)$, and those measured with respect to the axis of each jet are denoted by $(r_1,\theta_1)$ and $(r_2,\theta_2)$. Physical quantities are made dimensionless using $D$ and the free-stream sound speed $c_\infty$ and density $\rho_\infty$. Pressure is scaled with $\rho_\infty c^2 _\infty$ and temperature with $(\gamma -1)T_\infty$, where $\gamma$ is the ratio of specific heats. The jet Mach number is defined as $M_j = U_j / c_\infty$, with $U_j$ being the jet exit velocity. The jet temperature ratio is defined as $T_R = T_j / T_\infty$, where $T_j$ is the jet exit temperature and $T_\infty$ the free-stream temperature.

Let $\boldsymbol{q}'$ be a vector containing all the fluid variables of interest, e.g. the velocity $\boldsymbol{v}=(u,v,w)$, density $\rho$, pressure $p$ and temperature $T$. Linear instability theory (LST) studies small-amplitude disturbances superimposed on a time-invariant flow, either steady laminar or stationary turbulent mean flow, denoted here by $\bar{\boldsymbol{q}}$. Invoking the locally-parallel flow assumption, modal disturbances of the form
\begin{equation}\label{eqn:Methodology_Modal}
\boldsymbol{q}'(\boldsymbol{x},t) = \boldsymbol{q}(y,z)\, \ee^{\ii \left(k x - \omega t\right)} + \mathrm{c.c.},
\end{equation}
\noindent are introduced, where $\omega$ is the circular frequency, $k$ the streamwise wavenumber, $t$ the dimensionless time and c.c. denotes the complex conjugate. The Strouhal number defined as $St = f D/U_j$ is related to the dimensionless angular frequency by $\omega = 2 \pi M_j St$. The derivation of the LST problem continues by substituting the modal decomposition (\ref{eqn:Methodology_Modal}) in the linearised governing equations and recasting the result as an eigenvalue problem. The spatial instability framework is used in this work, which consists of prescribing a real frequency $\omega$ and obtaining the corresponding eigensolutions as the eigenvalue/eigenfunction pairs $(k,\boldsymbol{q})$. Thus, LST provides the dispersion relation $\mathcal{D}(\omega,k) = 0$ that governs the evolution of linear instability waves. 

LST is applied here to twin-jet configurations. As opposed to single round jets, axial symmetry cannot be exploited here to simplify the LST problem; however, the twin-jet mean flow is symmetric with respect to the $(x,y)$ and $(x,z)$ planes. Following \cite{Morris:JFM90}, the LST solutions are separated in four families corresponding to the even or odd character of their pressure field with respect to the two planes. The two-letter notation used by \citet{Rodriguez:CRM2018,Nogueira:JFM21} is adopted to classify the mode families, which is outlined in table \ref{tab:Methodology_families}. Two different formulations of the LST problem are used in this work, as described next.

\begin{table}
\caption{Classification of mode families depending on the symmetries. The fourth column shows the relation of the notation used here with that by \citet{Morris:JFM90}. The azimuthal dependence for each family is also shown. The last two columns show the values of $\phi_y$ and $\phi_z$ appearing in the vortex sheet model.}
\label{tab:Methodology_families}
\centering{
\begin{tabular*}{.85\textwidth}%
     {@{\extracolsep{\fill}}cccclccc}
\\
($x,y$)-plane	& ($x,z$)-plane	& Family 	& Alt. name &  Azimuthal modes & \hspace{5pt} & $\phi_y$ & $\phi_z$ \\ \hline
Even &	Even &	SS & I	 & $\cos(2n\theta)$	& & 1 & 1 \\
Odd  &  Even &  AS & II  & $\cos(2n+1)\theta$	& & -1 & -1 \\
Even &  Odd  &	SA & III & $\sin(2n+1)\theta$	& & -1 & 1 \\	
Odd  & 	Odd  &	AA & IV	 & $\sin(2n\theta)$	& & 1 & -1
\end{tabular*}
}
\end{table}

\subsection{Formulation 1: Complete compressible Navier-Stokes equations discretised in cartesian coordinates}
\label{sec:Method1}

This formulation is valid for finite-thickness jets of arbitrary shape and has been applied previously to twin round jets \citep{Rodriguez:CRM2018,Rodriguez:AIAA2021} and rectangular jets \citep{RodriguezPrasad:AIAA2021}.
The viscous compressible continuity, momentum and energy equations in cartesian coordinates are used as the departure point. Ideal gas is assumed and the impact of temperature on viscosity is neglected. In the linearisation, density and temperature are used for the mean flow and pressure and temperature for the disturbances. The modal form (\ref{eqn:Methodology_Modal}) is introduced and the resulting generalised eigenvalue problem is recast in the form
\begin{equation}\label{eqn:Methodology_EVP}
\mathsfbi{L} \boldsymbol{q} = k \mathsfbi{R} \boldsymbol{q},
\end{equation}
\noindent where matrix operators $\mathsfbi{L}$ and $\mathsfbi{R}$ depend on the mean flow and its derivatives, $\omega$, the physical parameters $Re$, $Ma$, $Pr$ and $\gamma$ and parameters describing the twin-jet configuration (e.g. $s/D$, $M_j$, $T_R$).


The rectangular domain $\Omega = [0,y_{\infty}]\times[0,z_{\infty}]$ is used in the discretisation. An analytical coordinate transformation is applied independently to the $y$ and $z$ coordinates to concentrate points around the jet shear layer. The odd/even behaviours of each family are imposed as boundary conditions along the $y = 0$ and $z = 0$ axes.
For modes that are respectively symmetric (S) and anti-symmetric (A) with respect to the $(x,y)-$plane, the conditions
\begin{eqnarray}
\mbox{S}: && \dfrac{\partial p}{\partial y} = \dfrac{\partial T}{\partial y} = \dfrac{\partial u}{\partial y} = \dfrac{\partial v}{\partial y} = 0, \quad w = 0, \\
\mbox{A}: && p = T = u = v = 0, \quad \dfrac{\partial w}{\partial y} = 0,
\end{eqnarray}
\noindent are imposed at $y=0$, and similarly for the behaviour with respect to the $(x,z)-$plane. Vanishing of the disturbance velocity and temperature is imposed as far-field boundary conditions. A Neumann condition is imposed for the pressure. The rectangular domain $\Omega =[0,7.5]\times[0,5]$ was checked to be large enough for the convergence of the results for the case with the largest jet separation considered herein ($s/D = 5$), and is used for all calculations.

The linear operators $\mathsfbi{L}$ and $\mathsfbi{R}$ are discretised using a combination of variable-stencil high-order finite differences and sparse algebra that exploits the banded structure of the differentiation matrices. A 7-point stencil is used, which results in the optimal balance between accuracy and computational cost \citep{Gennaro:AIAAJ13,Rodriguez:ICOSAHOM16}. After discretisation of the linear operators, the matrix eigenvalue problem (\ref{eqn:Methodology_EVP}) is solved using an in-house sparse implementation of the shift-and-invert Arnoldi's algorithm \citep{Arnoldi}. Arnoldi's algorithm requires the solution of a number of linear problems, which is accomplished using the package MUMPS \citep{Amestoy2001}.

This formulation can also be applied without exploiting/imposing the symmetries, as done in \cite{Rodriguez:CRM2018}. In this case, the computational domain used is $\Omega = [-y_{\infty},y_{\infty}]\times[-z_{\infty},z_{\infty}]$ and the same mode families are recovered. The results of the present formulation have been cross-validated with the formulation presented by \cite{Nogueira:JFM21}, that employs a Floquet formalism on the azimuthal direction to reduce the computation to a sector of the azimuthal domain and discretises it using a two-dimensional mesh in polar coordinates.

\subsection{Formulation 2: Vortex-sheet formulation for twin jets}
\label{sec:Method2}

The vortex sheet model treats the shear layer as a boundary of infinitesimal width with a uniform streamwise velocity inside the jet and zero streamwise velocity outside \citep{Lessen:JFM65,Michalke:DLR1970}. The inviscid equations governing the linear instability waves, upon the introduction of the modal form (\ref{eqn:Methodology_Modal}) written in terms of the cylindrical coordindates, reduce to the Helmholtz equation for the disturbance pressure $p$:
\begin{equation}
\dfrac{\partial^2 p}{\partial r^2} + \dfrac{1}{r}\dfrac{\partial p}{\partial r} + \dfrac{1}{r^2}\dfrac{\partial^2 p}{\partial \theta^2} - \lambda^2 p = 0. 
\end{equation}
\noindent  This equation describes the disturbance pressure in the inner ($i$) and outer ($o$) regions to the vortex sheet, depending on the definition of $\lambda$:
\begin{equation}
\lambda_{i} = \sqrt{k^2 -\frac{1}{T_R}(\omega-M_{j} k)^2}, \quad
\lambda_{o} = \sqrt{k^2 -\omega^2}.
\label{eqn:Lambda}
\end{equation}

In the twin-jet configuration \citep{Morris:JFM90,DU1993,Stavropoulos:AIAA2021}, the inner and outer solutions are written in forms consistent with the separation in even/odd families as
\begin{eqnarray}
p_{i,1} (r_1,\theta_1) & = & \sum \limits^\infty _{m=0} \hat{A}_m I_m(\lambda_i r_1) \cos(m\theta_1) + \hat{B}_m I_m(\lambda_i r_1) \sin(m \theta_1), \label{eqn:VS_innersol} \\
p_o(r_1,\theta_1,r_2,\theta_2) & = & 
   \sum \limits^\infty _{m=0} A_m \left[ K_m(\lambda_o r_1) \cos(m\theta_1) + (-1)^m K_m(\lambda_o r_2) \cos(m\theta_2)\right] + \nonumber \\
&& \sum \limits^\infty _{m=0} B_m \left[ K_m(\lambda_o r_1) \cos(m\theta_1) - (-1)^m K_m(\lambda_o r_2) \cos(m\theta_2)\right] + \nonumber \\
&& \sum \limits^\infty _{m=0} C_m \left[ K_m(\lambda_o r_1) \sin(m\theta_1) + (-1)^m K_m(\lambda_o r_2) \sin(m\theta_2)\right] + \nonumber \\
&& \sum \limits^\infty _{m=0} D_m \left[ K_m(\lambda_o r_1) \sin(m\theta_1) - (-1)^m K_m(\lambda_o r_2) \sin(m\theta_2)\right] \label{eqn:VS_outersol}
\end{eqnarray}
\noindent where $I_m$ and $K_m$ are the modified Bessel functions of first and second kind. Each line of (\ref{eqn:VS_outersol}) corresponds to one of the four families. 

The inner and outer solutions are matched at the ideally-expanded diameter $D_j$, that is imposed to be equal to $D$ in this work. Matching conditions impose the continuity of the pressure and displacement across the vortex sheet:
\begin{equation}\label{eqn:matching1}
p_i(\lambda_i D_j / 2) = p_o (\lambda_o D_j / 2),
\end{equation}
\begin{equation}\label{eqn:matching2}
\left. \dfrac{\partial p_i}{\partial r} \right|_{r=D_j/2} = \dfrac{1}{T_R}\dfrac{(\omega - kM_j)^2}{\omega^2}\left. \dfrac{\partial p_o}{\partial r} \right|_{r=D_j/2}.
\end{equation}

In order to impose the matching conditions, the outer solution (\ref{eqn:VS_outersol}) is re-written in terms of the coordinates of a single jet using Graf's addition theorem \citep{Abramowitz:1964}:
\begin{eqnarray}
K_m (\lambda_o r_2) \cos(m\theta_2) & =&  \sum \limits^{\infty}_{n = -\infty} (-1)^n K_{m-n} (\lambda_o s)I_n(\lambda_o r_1) \cos(n \theta_1), \label{eqn:Graf1}\\
K_m (\lambda_o r_2) \sin(m\theta_2) & =&  \sum \limits^{\infty}_{n = -\infty} (-1)^n K_{m-n} (\lambda_o s)I_n(\lambda_o r_1) \sin(n \theta_1). \label{eqn:Graf2}
\end{eqnarray}

Combining equations (\ref{eqn:VS_innersol})-(\ref{eqn:Graf2}) and collecting terms corresponding to the same family yields the dispersion relation for a twin-jet system as

\begin{equation}
\mathcal{D}(\omega,k;M_j,s/D,D_j) = \sum\limits_{m=0}^{\infty}\psi_{m}[a_{nn}\delta_{mn} + \phi_y (-1)^m c_{mn}] = 0,
\label{eqn:Disp}
\end{equation}
\noindent where $\psi_m$ correspond to the coeffients $A_m, B_m, C_m$ or $D_m$ depending on the solution family, $\delta_{mn}$ is the Kronecker delta and
\begin{equation}
a_{nn} = \frac{1}{\left(1-\frac{kM_{j}}{\omega}\right)^2}-\frac{1}{T_R}\frac{\lambda_o}{\lambda_i}\frac{K_n^{'}\left(\frac{D_{j}\lambda_o}{2}\right) I_n\left(\frac{D_{j}\lambda_i}{2}\right)}{I_n^{'}\left(\frac{D_{j}\lambda_i}{2}\right)K_n\left(\frac{D_{j}\lambda_o}{2}\right)},
\label{eqn:Ann}
\end{equation}
\begin{eqnarray}
c_{mn} & = & (-1)^n\epsilon_n[K_{m-n}(\lambda_o s) + \phi_z
K_{m+n}(\lambda_0s)] \nonumber \\
&& \times \left[\frac{I_n\left(\frac{D_{j}\lambda_o}{2}\right)}{K_n\left(\frac{D_{j}\lambda_0}{2}\right)}\frac{1}{\left(1-\frac{kM_{j}}{\omega}\right)^2}-\frac{1}{T_R}\frac{\lambda_o}{\lambda_i}\frac{I_n\left(\frac{D_{j}\lambda_i}{2}\right)I_n^{'}\left(\frac{D_{j}\lambda_o}{2}\right)}{K_n\left(\frac{D_{j}\lambda_o}{2}\right)I_n^{'}\left(\frac{D_{j}\lambda_i}{2}\right)}\right].
\label{eqn:Cmn}
\end{eqnarray}

Here, $\epsilon_{n}$ is equal to 0.5 if $n$ = 0 and 1 otherwise. The mode families are imposed through the factors $\phi_y$ and $\phi_z$ in equations (\ref{eqn:Disp}) and (\ref{eqn:Cmn}) (see table \ref{tab:Methodology_families}). For large $s$ the first factor in brackets in equation (\ref{eqn:Cmn}) tends to zero leaving only those from equation (\ref{eqn:Ann}), which recovers the dispersion relation for a single jet \citep{Lessen:JFM65,Towne:JFM2017}. As opposed to the case of a single jet, equation (\ref{eqn:Disp}) couples all the azimuthal $m$ numbers. For the practical solution, the summatory is truncated at a finite $N$. The results presented in this work are computed using $N = 5$, but their convergence has been checked with a maximum $N$ of 10.


\section{Results}
\label{sec:Results}

\subsection{Finite-thickness single and twin-jet mean flows}
\label{sec:MeanFlow}

The calculations of finite-thickness single and twin jets in this work employ analytic mean flows for the sake of reproducibility. The analytical velocity profile proposed by \cite{Michalke:DLR1970,MichalkePrAS1984} is used:
\begin{equation}\label{eqn:Results_BF}
\frac{\bar{u}}{U_j} = \dfrac{1}{2} \left[ 1 + \tanh\left(  \frac{1}{4}\frac{R}{\theta} \left(\frac{R}{r} - \frac{r}{R} \right)  \right) \right],
\end{equation}
\noindent where $r$ is the radial coordinate measured from the jet centre for a single jet and $R$ is the nozzle radius, $R = D/2$.
A tailored mean flow field is constructed for twin jet configurations by combining the flow fields corresponding to two isolated round jets of the form (\ref{eqn:Results_BF}) aligned with the $x-$direction and with centres placed symmetrically at the coordinates $(y,z)=(\pm s/2,0)$, as shown in figure \ref{fig:sketch}. 
This is a good approximation of the twin-jet mean flow in the region immediate downstream of the nozzles, known as the converging region \citep{Okamoto:JSME1985,Moustafa:AIAAJ1994}, where the shear layers of the two jets are still well separated, and has been used in the past in the linear stability calculations by \cite{Morris:JFM90,Rodriguez:CRM2018,Rodriguez:AIAA2021}.
A zero mean pressure gradient is assumed and the Crocco-Busemann relation particularised for the temperature ratio $T_R$, 
\begin{equation}
\frac{\bar{T}}{T_\infty} = \left(U_j - \bar{u}  \right) \frac{\bar{u}}{2} + \frac{1}{\gamma -1}\left(1 + (T_R-1)\dfrac{\bar{u}}{U_j} \right),
\end{equation}
\noindent is used to determine the mean temperature $\bar{T}$. All the quantities in this expression are dimensionless, as explained in section \ref{sec:Methodology}. The mean density field $\bar{\rho}$ is obtained from the state equation. Following the parallel-flow assumption, cross-stream mean velocity components are neglected. For the finite-thickness calculations herein, the jet Mach number is set at $M_j = 1.5$ and $T_R = 1$ and the parameter $R/\theta = 12.5$. This value of $R/\theta$ is representative of the thin shear layer in the first diameter from the nozzle \citep{CrightonGaster:JFM76,MichalkePrAS1984}. The Reynolds number $Re = 5\times 10^4$ is used in the calculations but identical results are recovered for $Re = 10^5$, showing that the effect of viscosity is negligible.

\subsection{The eigenspectra of single and twin jets}
\label{sec:Results1}

\begin{figure}
\centerline{\begin{tabular}{cc}
&
\includegraphics[trim = 0 87mm 0 7mm, clip, width=.45\textwidth]{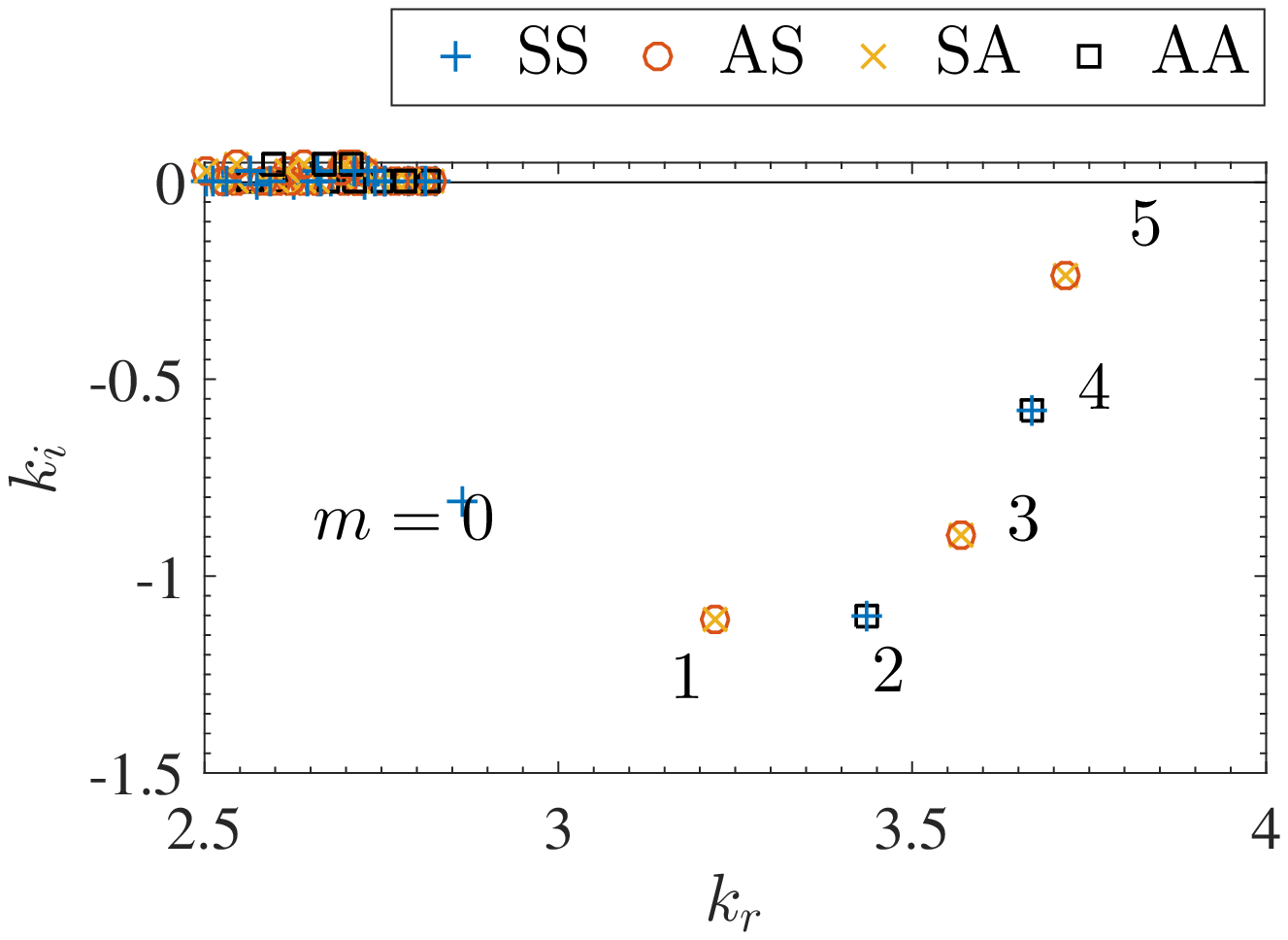} \\
\includegraphics[width = .45\textwidth]{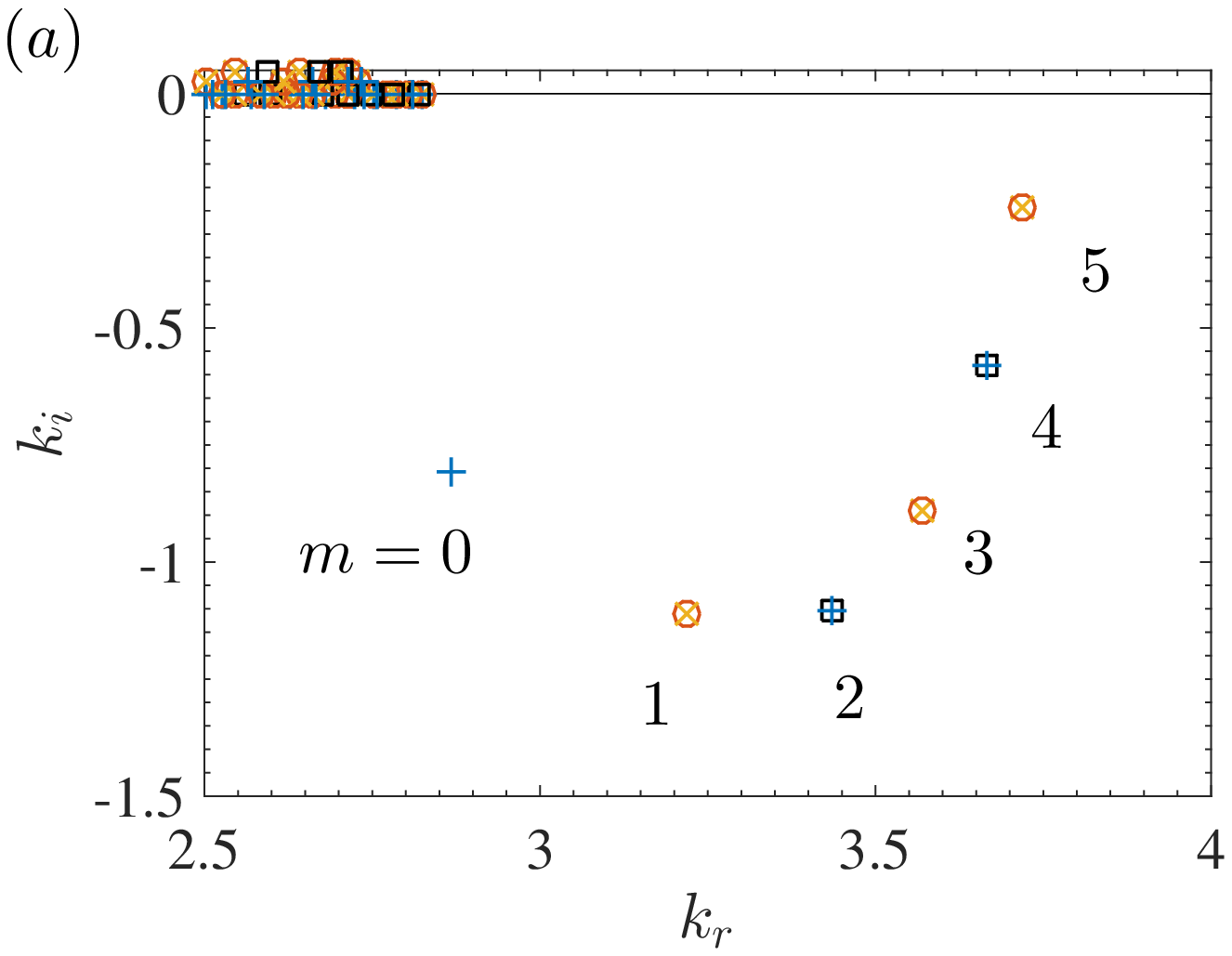} &
\includegraphics[width = .45\textwidth]{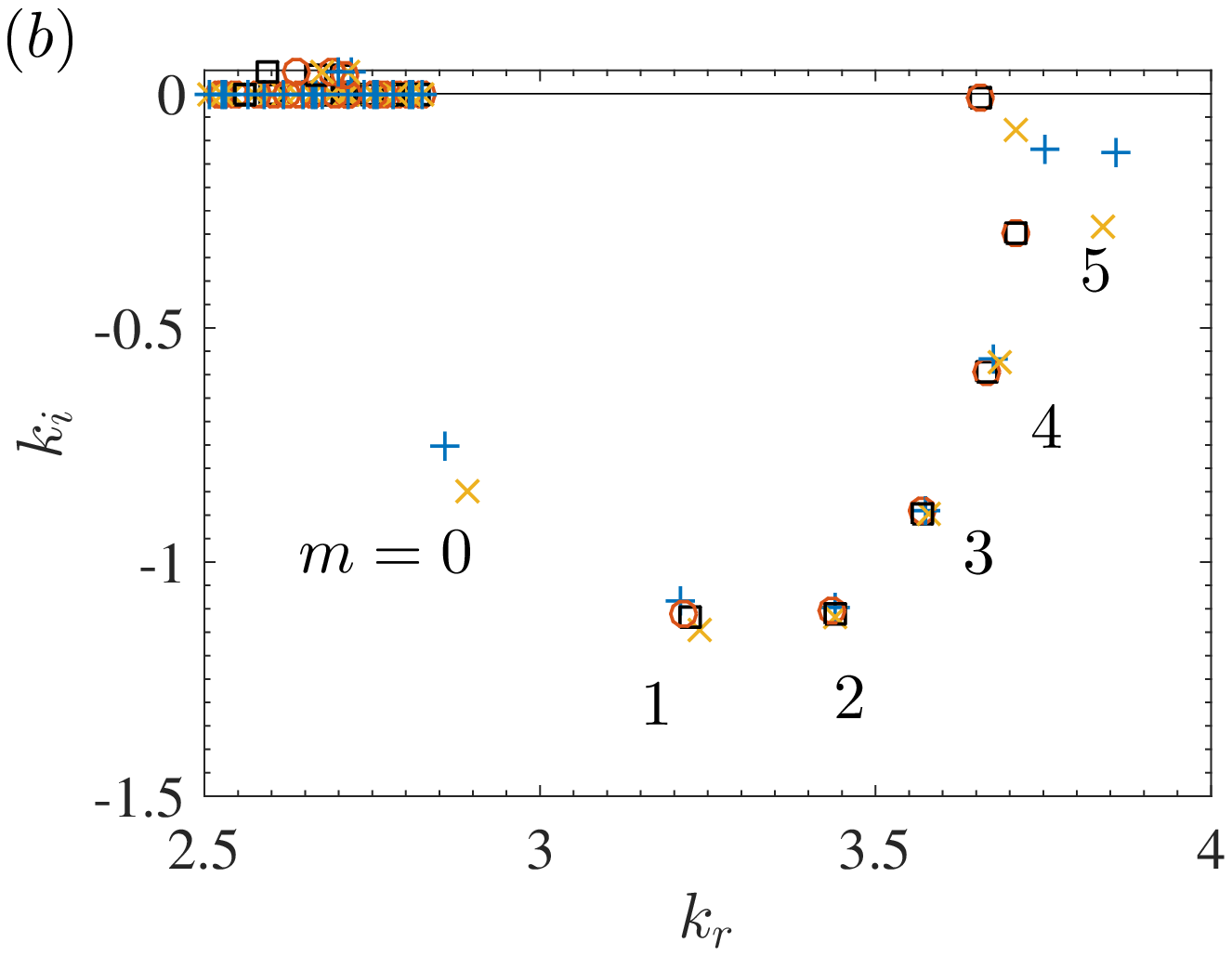} \\
\end{tabular}}
\caption{LST eigenspectra corresponding to $M_j=1.5, T_R = 1, R/\theta = 12.5$ and $St=0.3$ and the four solution families: SS ($+$), AS ($\circ$), SA ($\times$), AA ($\square$). ($a$) Single jet; ($b$) Twin jet with separation $s/D = 2.2$.}
\label{fig:EigVals}
\end{figure}

\begin{figure}
\centerline{\begin{tabular}{cc}
\includegraphics[trim=0cm 1.5cm 0cm 2.5cm, clip=true, width=.55\textwidth]{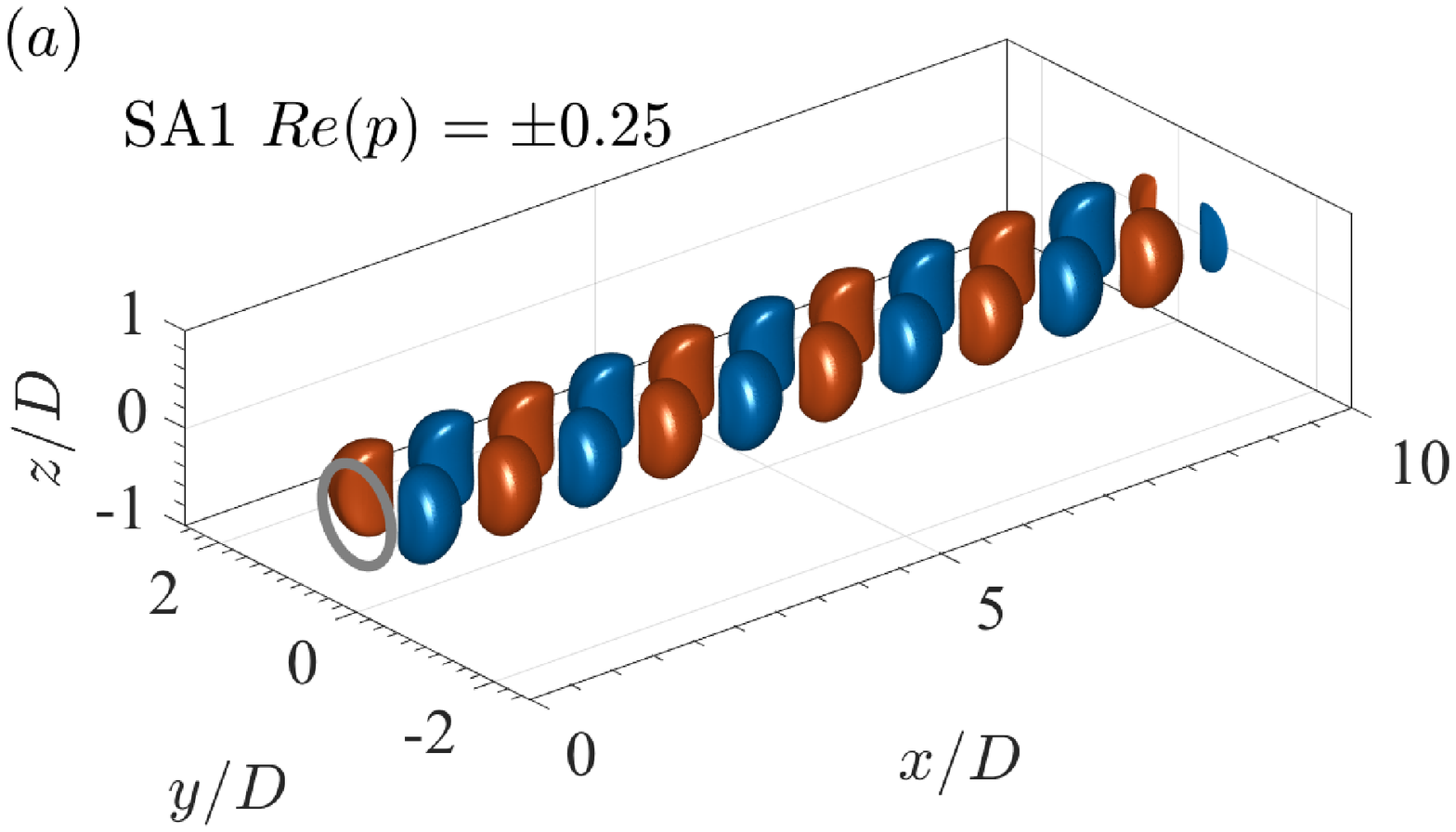} & 
\includegraphics[width=.45\textwidth]{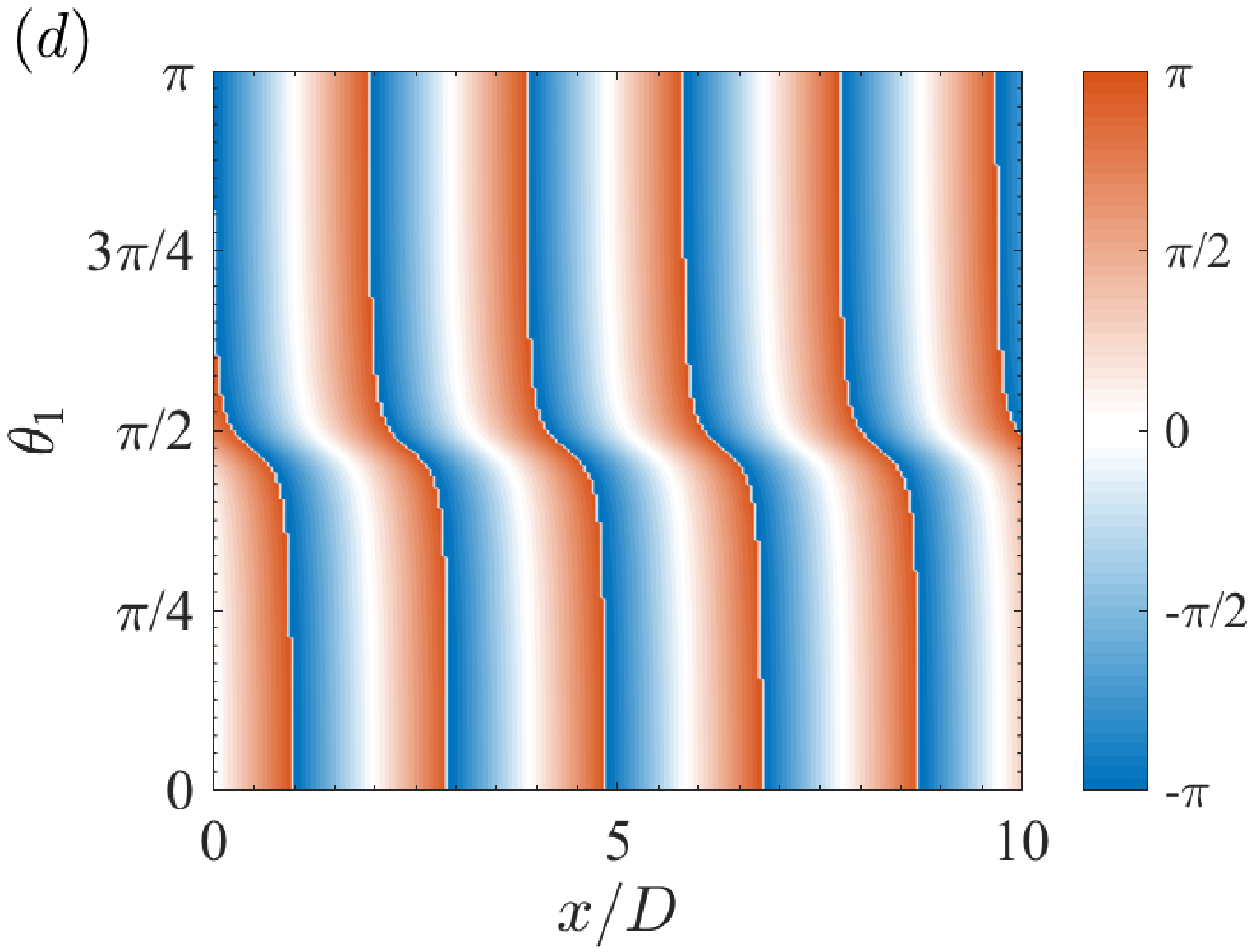} \\
\includegraphics[trim=0cm 1.5cm 0cm 2.5cm, clip=true, width=.55\textwidth]{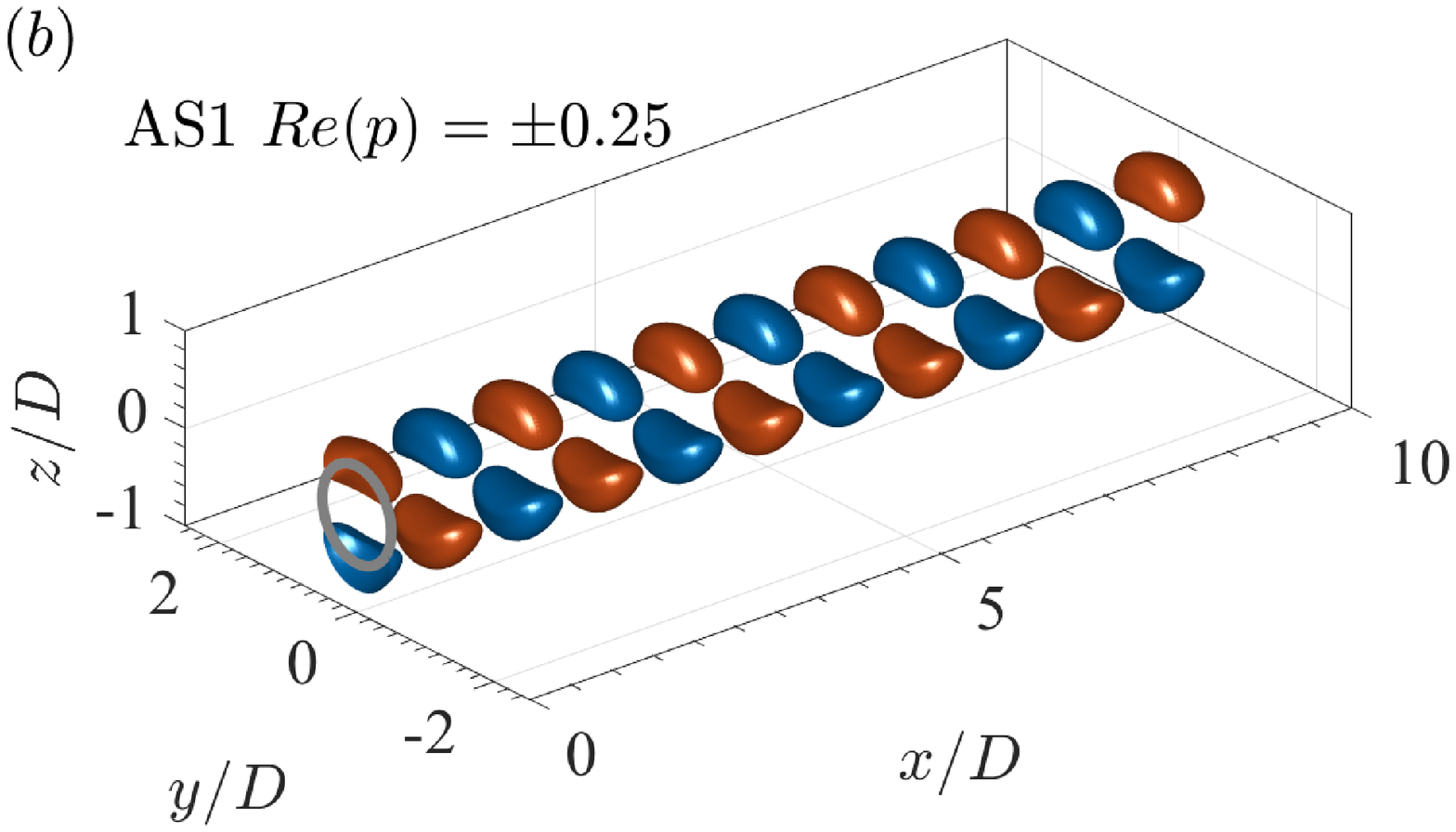} & 
\includegraphics[width=.45\textwidth]{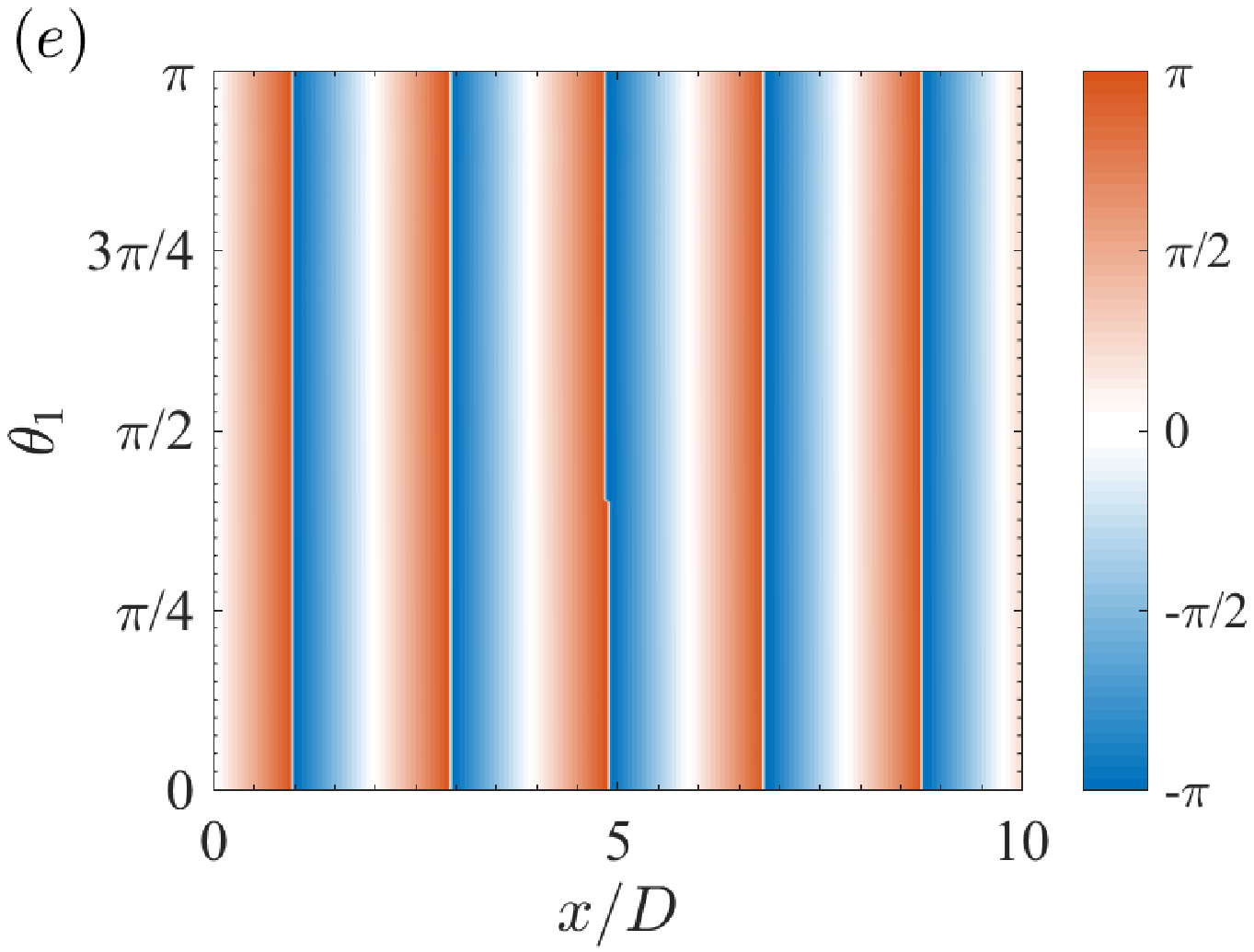} \\
\includegraphics[trim=0cm 1.5cm 0cm 2.5cm, clip=true, width=.55\textwidth]{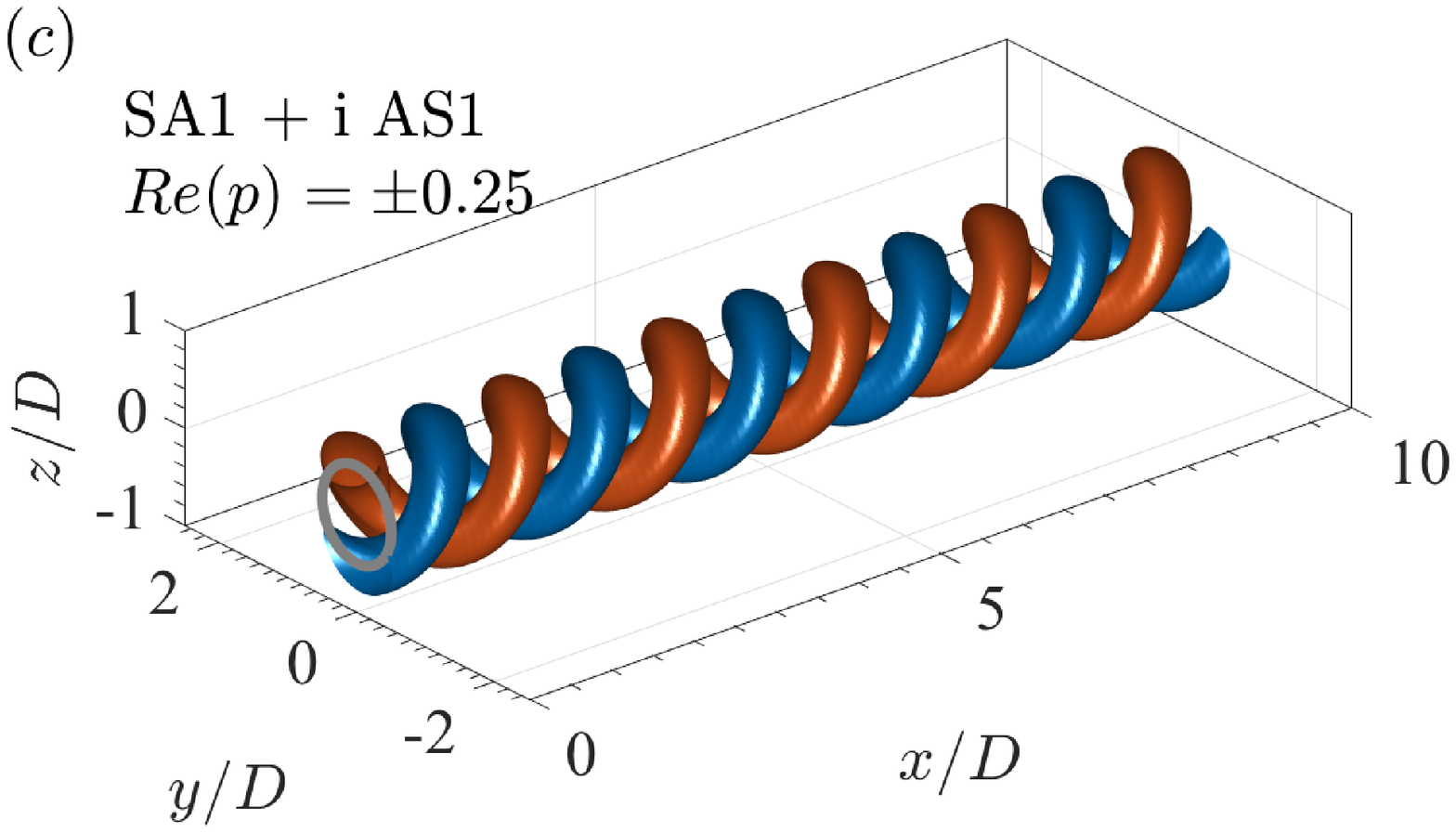} & 
\includegraphics[width=.45\textwidth]{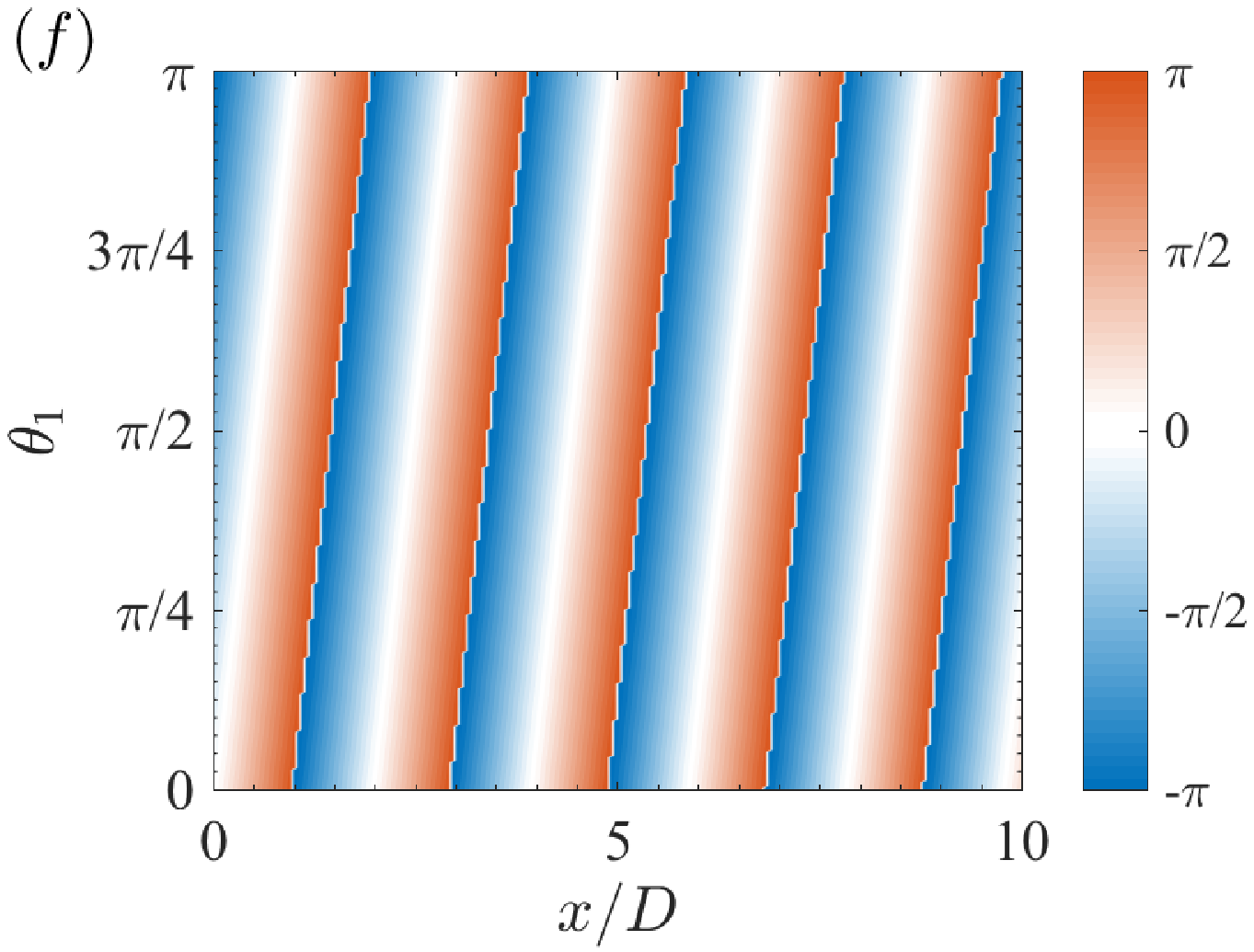} \\
\end{tabular}}
\caption{Pressure eigenfunctions corresponding to a single jet at $M_j=1.5, T_R = 1, R/\theta = 12.5$ and $St = 0.3$. The streamwise dependence is obtained from the corresponding eigenvalue, eliminating the spatial growth. Left column: iso-contours of the real pressure component. The grey circle shows the nozzle circumference. Right column: phase angle $\phi$ at a cylinder of radius 0.75$D$. ($a,d$) SA1; ($b,e$) AS1; ($c,f$) Linear combination of SA1 and AS1 to produce the $m=1$ helical mode. }
\label{fig:EigVec_SJ_helical}
\end{figure}

This section discusses the results of the LST analysis based on cartesian coordinates, described in section \ref{sec:Method2}, when applied to single and twin jets.

\subsubsection{Single round jets}

When the local stability analysis is applied to single round jets, the axial symmetry of the mean flow allows for the introduction of azimuthal Fourier modes of the form $\exp(\ii m\theta)$ \citep[e.g.][]{Gill:PF65,MichalkePrAS1984}. This is usually exploited to reduce the eigenvalue problem to a one-dimensional one, dependent on the radial coordinate alone. For each $m$, the stability analysis recovers different families of eigensolutions \citep{TamHuJFM1989,Rodriguez:PF13}, but only one discrete eigenmode is associated with the Kelvin-Helmholtz (K-H) instability.

The present approach is based on Cartesian coordinates, discretising only the first quadrant of the $(y,z)-$plane and imposing symmetric/anti-symmetric conditions at the $y= 0$ and $z= 0$ planes. As a consequence, it does not isolate the individual azimuthal eigenmodes. Conversely, the imposed symmetries separate the eigenmodes in four families: SS, AS, SA and AA, as shown in table \ref{tab:Methodology_families} and discussed elsewhere \citep{Morris:JFM90,Rodriguez:AIAA2021,RodriguezPrasad:AIAA2021}.
When this approach is applied to a single jet centred at the origin of coordinates, each eigenspectrum contains a number of discrete K-H eigenmodes, each one corresponding to one $m$. Figure \ref{fig:EigVals}($a$) shows the eigenspectra for a single jet at $St=0.3$. The different symbols correspond to one of the solution families. Except for $m=0$, which is only recovered in family SS, all other eigenvalues appear repeated in two families. In particular, the eigenmodes corresponding to $m=1$ are recovered in families AS and SA; in the following these eigenmodes are referred to as AS1 and SA1, respectively.
Their pressure eigenfunctions are shown in figure \ref{fig:EigVec_SJ_helical}($a$,$b$). The three-dimensional pressure fields are reconstructed following (\ref{eqn:Methodology_Modal}), with the eigenfunction $p(y,z)$ normalised. 
To aid in the visualisation, an arbitrary streamwise domain of length $10D$ is used and the pressure field amplitude is multiplied by $\exp(-k_i x)$ to cancel the spatial growth, so that the amplitude remains constant. 
The pressure vanishes simultaneously at one of the coordinate axes (different for SA and AS) for the real and imaginary components. This behaviour is enforced here by the symmetry/anti-symmetry conditions for each family, but a computation using the domain $\Omega = [-y_{\infty},y_{\infty}]\times[-z_{\infty},z_{\infty}]$ and not imposing the symmetries delivers identical families of eigenfunctions \citep{Rodriguez:CRM2018}. Figure \ref{fig:EigVec_SJ_helical}($d$,$e$) shows the phase of the pressure field,  $\phi = \arctan(\Imag(p)/\Real(p))$,  extracted at a cylinder of radius 0.75$D$ aligned with the jet axis. The azimuthal angle $\theta_1$ is measured from the positive $y$ axis as shown in figure \ref{fig:sketch}. Only the subdomain $\theta_1 \in [0,\pi]$ is shown, but the omitted region can be inferred from the symmetries. Mode SA1 presents a constant phase for $-\pi/2 < \theta_1 < \pi/2$ with is shifted an angle $\pi$ for other values of $\theta_1$. Mode AS1 shows a similar behaviour but with the phase shift at $\theta_1 = \pi$. Each of these eigenfunctions describes a jet flapping motion.
However, because the eigenvalues corresponding to the modes AS1 and SA1 are identical ($k_{SA1} = k_{AS1}$), a linear combination of them is also a valid solution to the general problem. With appropriately chosen amplitude coefficients for each eigenmode, their linear combination recovers the features of an helical mode. This is illustrated in figure \ref{fig:EigVec_SJ_helical}($c,f$), in which the pressure fields of modes AS1 and SA1 are added as

\begin{equation}\label{eqn:Results_SJ_helical}
p(x,y,z,t) = \hat{p}_{\mathrm{SA1}} (y,z)\, \ee^{\ii(k_{\mathrm{SA1}} x - \omega t)} + \ii\ \hat{p}_{\mathrm{AS1}} (y,z)\, \ee^{ \ii(k_{\mathrm{AS1}} x - \omega t) } + \mathrm{c.c.}.
\end{equation}

\noindent The helical nature of the resulting mode is manisfest in the linear dependence of the phase $\phi$ on the azimuthal angle $\theta_1$.

\subsubsection{Twin round jets}

\begin{figure}
\centerline{\begin{tabular}{cc}
&
\includegraphics[trim = 0 87mm 0 7mm, clip, width=.45\textwidth]{legends} \\
\includegraphics[width = .45\textwidth]{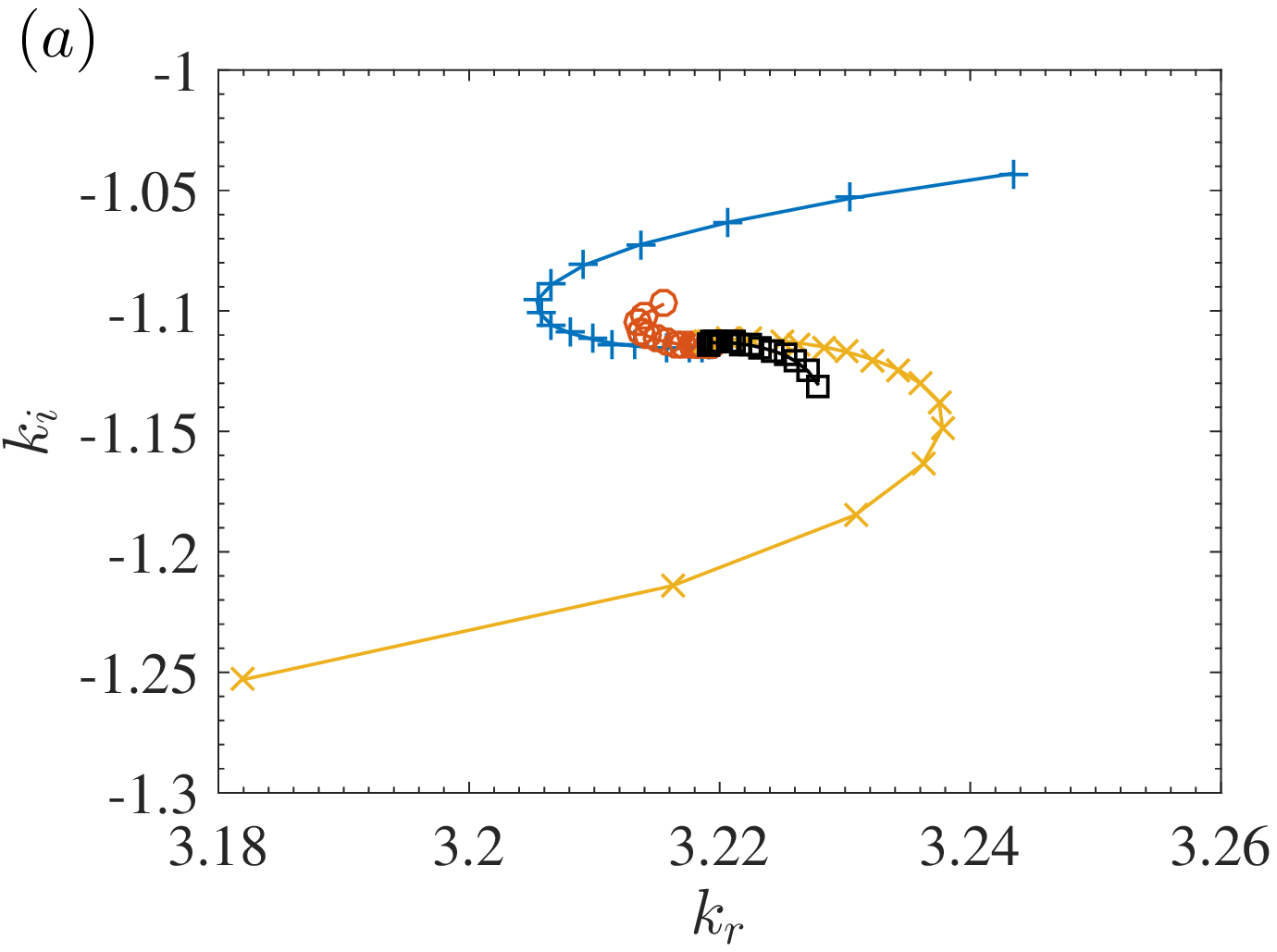} &
\includegraphics[width = .45\textwidth]{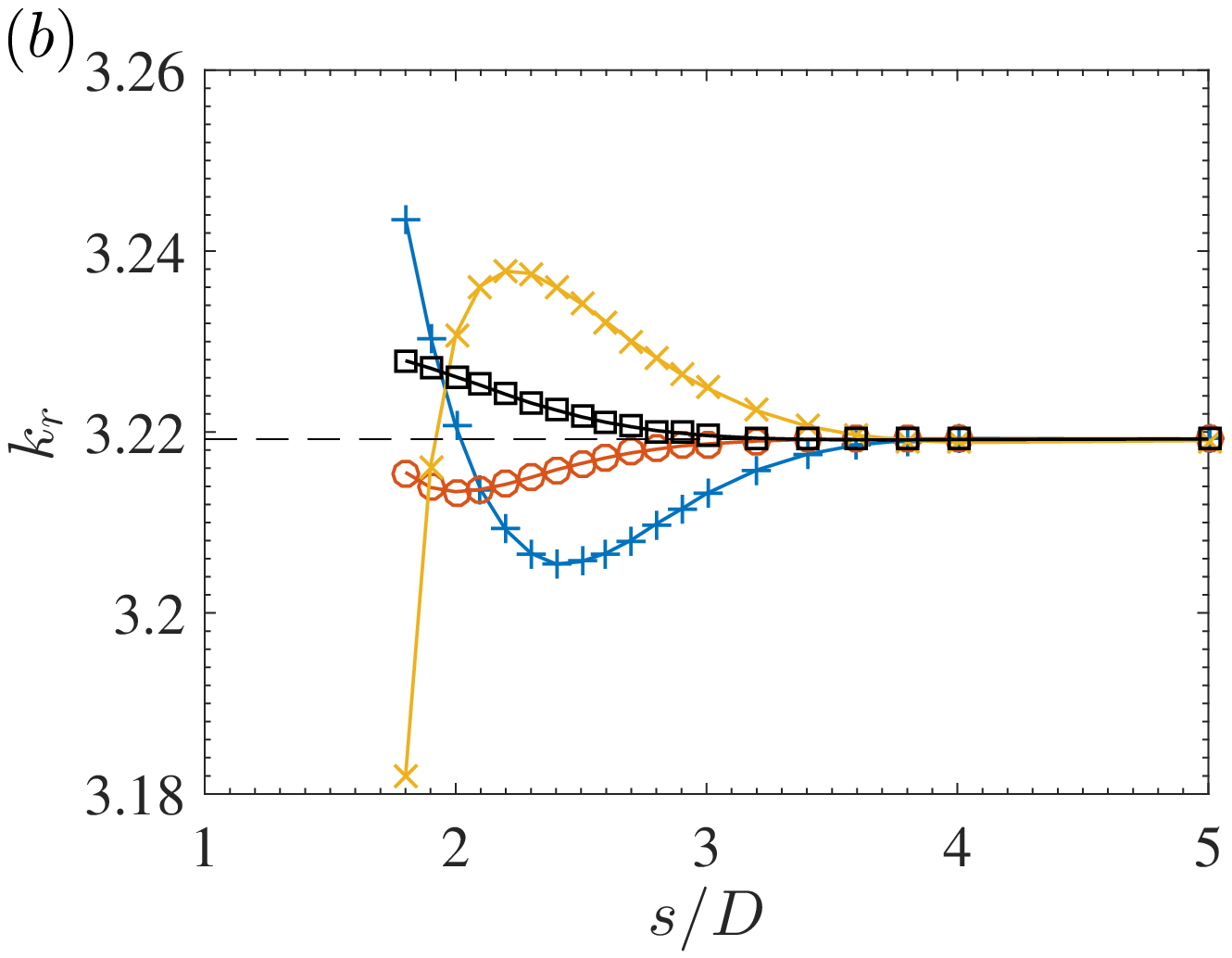} \\
\includegraphics[width = .45\textwidth]{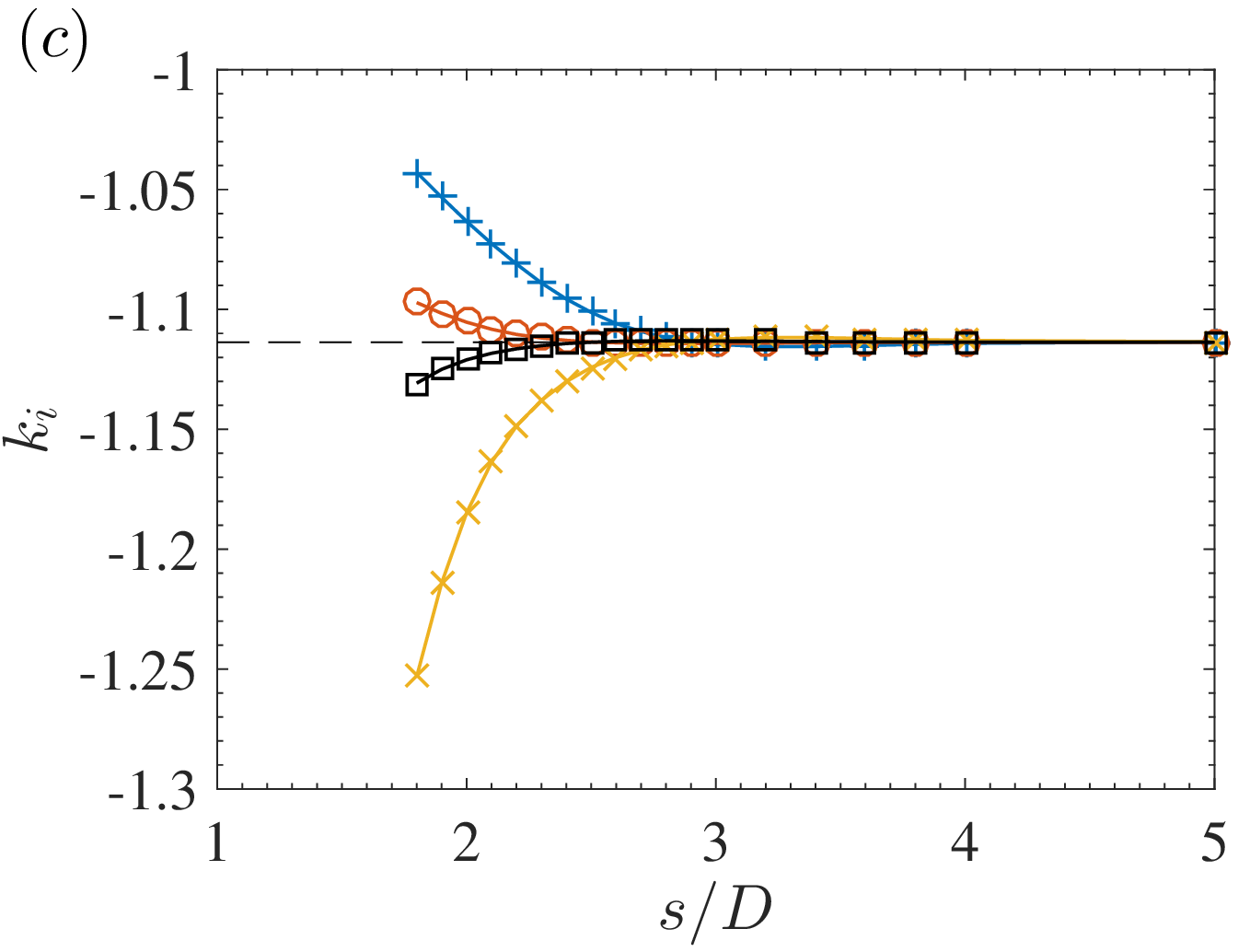} &
\includegraphics[width = .45\textwidth]{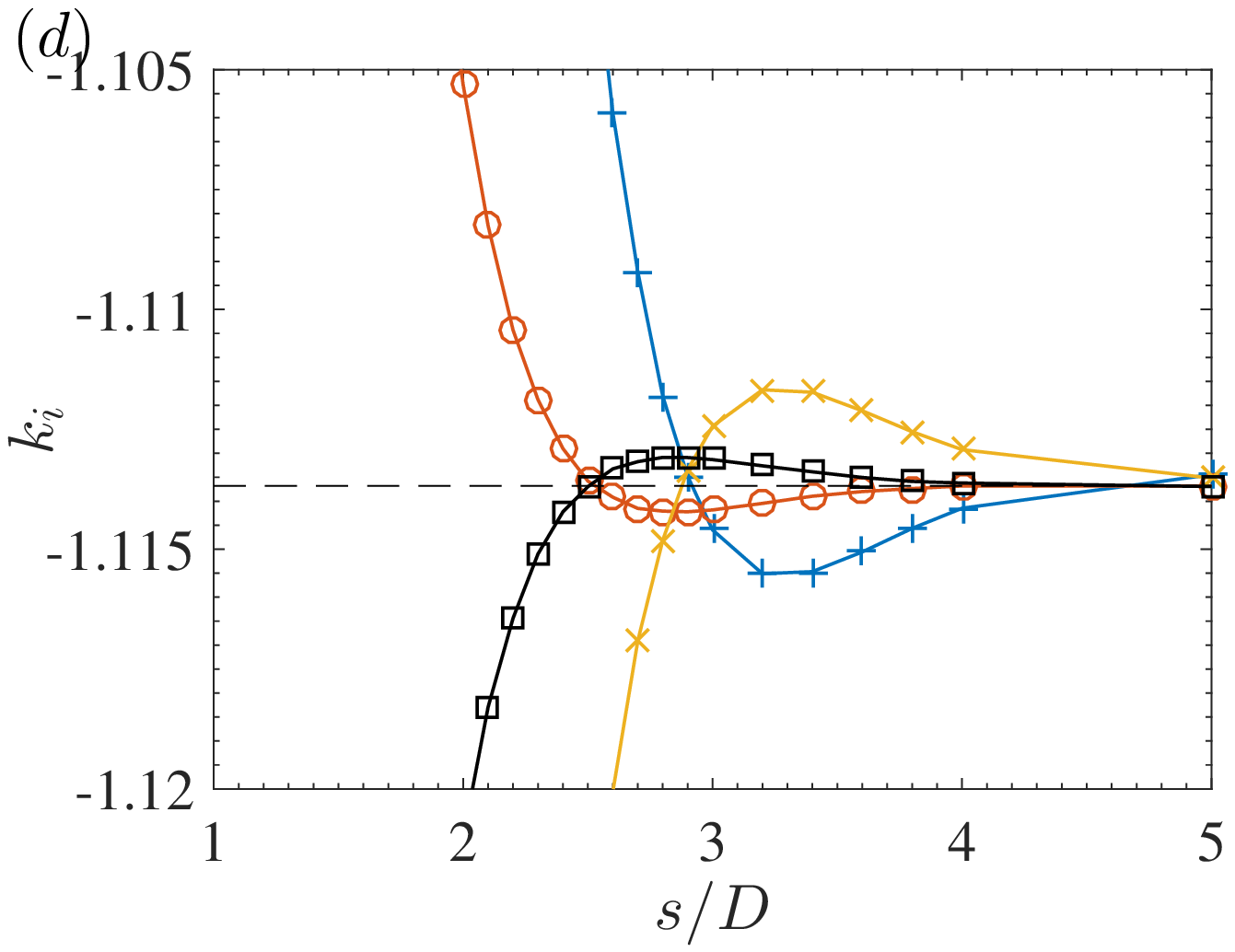} \\
\end{tabular}}
\caption{Dependence of the $m=1$ eigenvalues on the jet separation $s/D$: ($a$) In the complex $k$ plane. The eigenvalues spiral outwards with $s/D$ decreasing from 5 to 1.8. ($b$) Real and ($c,d$) imaginary parts. Panel ($d$) is a zoom in of panel ($c$). $M_j=1.5, T_R = 1, R/\theta = 12.5$ and $St= 0.3$. SS ($+$), AS ($\circ$), SA ($\times$), AA ($\square$). The horizontal dashed line corresponds to the $m=1$ mode of the single jet.}
\label{fig:EigVal_m1}
\end{figure}

 For the twin-jet configuration, the jet axes are not located at the origin of coordinates; while the nomenclature used for the single jet remains valid, the four solution families do not separate the odd and even $m$ modes. Instead, each family contains one eigenmode for each of the $m > 0$ modes. The symmetry of the mean flow about the mid-plane leads to the recovery of the $m=0$ mode only in the SS and SA families. Representative eigenspectra for a twin jet configuration with separation $s/D = 2.2$ at $St=0.3$ is shown in figure \ref{fig:EigVals}($b$), to be compared with the single jet eigenspectra in figure \ref{fig:EigVals}($a$).

The interaction between the fluctuation fields of the two jets breaks the azimuthal symmetry of the eigenmodes and the same $m$ mode presents different eigenvalues $k$, one for each symmetry family. This shift of the eigenvalues with respect to the single jet ones is stronger for $m=0$ and gradually reduces for increasing $m$. This behaviour is explained by the asymptotic decay of the single-jet eigenfunctions as $r\to \infty$, $\boldsymbol{q} \sim r^{-1/2} \exp\left(-r \sqrt{k^2 - \omega^2} \right)$ \citep{Abramowitz:1964}. For a fixed frequency $\omega$, the axial wavenumber $k$ increases with increasing $m$, and hence leads to a faster radial decay of the fluctuations. Consequently, the amplitude of the pressure field associated with one jet that reaches the other jet is reduced. For the same reason, the magnitude of the eigenvalue shift is also inversely proportional to the jet separation $s/D$, as shown next.

In what follows attention is focused on the $m=1$ modes, which present the largest growth rates for most of the conditions considered. These modes are denoted by SS1, SA1, AS1 and AA1, where the two letters correspond to the symmetry family and the number to the corresponding azimuthal number. Figure \ref{fig:EigVal_m1} shows the eigenvalues corresponding to each of the four families as a function of the jet separation. While the magnitude of the eigenvalue shift grows continuously with decreasing $s/D$, the dependence of the real and imaginary parts of the eigenvalues is non-monotonic and different for each family. 
The shift is stronger for modes SS1 and SA1, which correspond respectivaly to in-phase (varicose) and counter-phase (sinuous) lateral flapping oscillations in the plane containing the jet axes. For jet separation above $s/D \approx 2$, the real part of the wavenumber $k$ is reduced (equivalently, the phase velocity is increased) for SA1, while it is increased for SS1. This behaviour is reversed for very small jet separations. The flapping modes in the vertical (perpendicular) plane follow a similar trend: the in-phase mode AS1 reduces its phase velocity resulting from the presence of the other jet and the counter-phase AA1 increases it. 
As shown in figure \ref{fig:EigVal_m1}($c$) and more clearly in the zoomed-in version \ref{fig:EigVal_m1}($d$), the spatial growth rate presents a similar behaviour for the two pairs of modes: 

(i) The in-phase flapping modes SS1 (blue) and AS1 (red) exhibit an increase of the growth rate ($-k_i$) as $s/D$ is reduced from $s/D \rightarrow \infty$, due to jet interactions. For decreasing separations, the growth rate eventually reaches a maximum, then equals that of the single jet and then becomes smaller, i.e. close jet proximity stabilises the in-phase flapping modes. The jet separations for the destabilisation/stabilisation inversion are different for the two modes: $s/D \approx 2.8$ for SS1 and 2.5 for AS1. The variation of the growth rate is of bigger amplitude for the varicose lateral flapping mode SS1.

(ii) The counter-phase modes SA1 (yellow) and AA1 (black) present opposite trends to the in-phase modes. These modes have the largest growth rates for small jet separations and, in particular, mode SA1 corresponding to sinuous lateral flapping is the most amplified one.

\begin{figure}
\centerline{\begin{tabular}{cc}
\includegraphics[trim=0cm 1.5cm 0cm 2.5cm, clip=true, width=.55\textwidth]{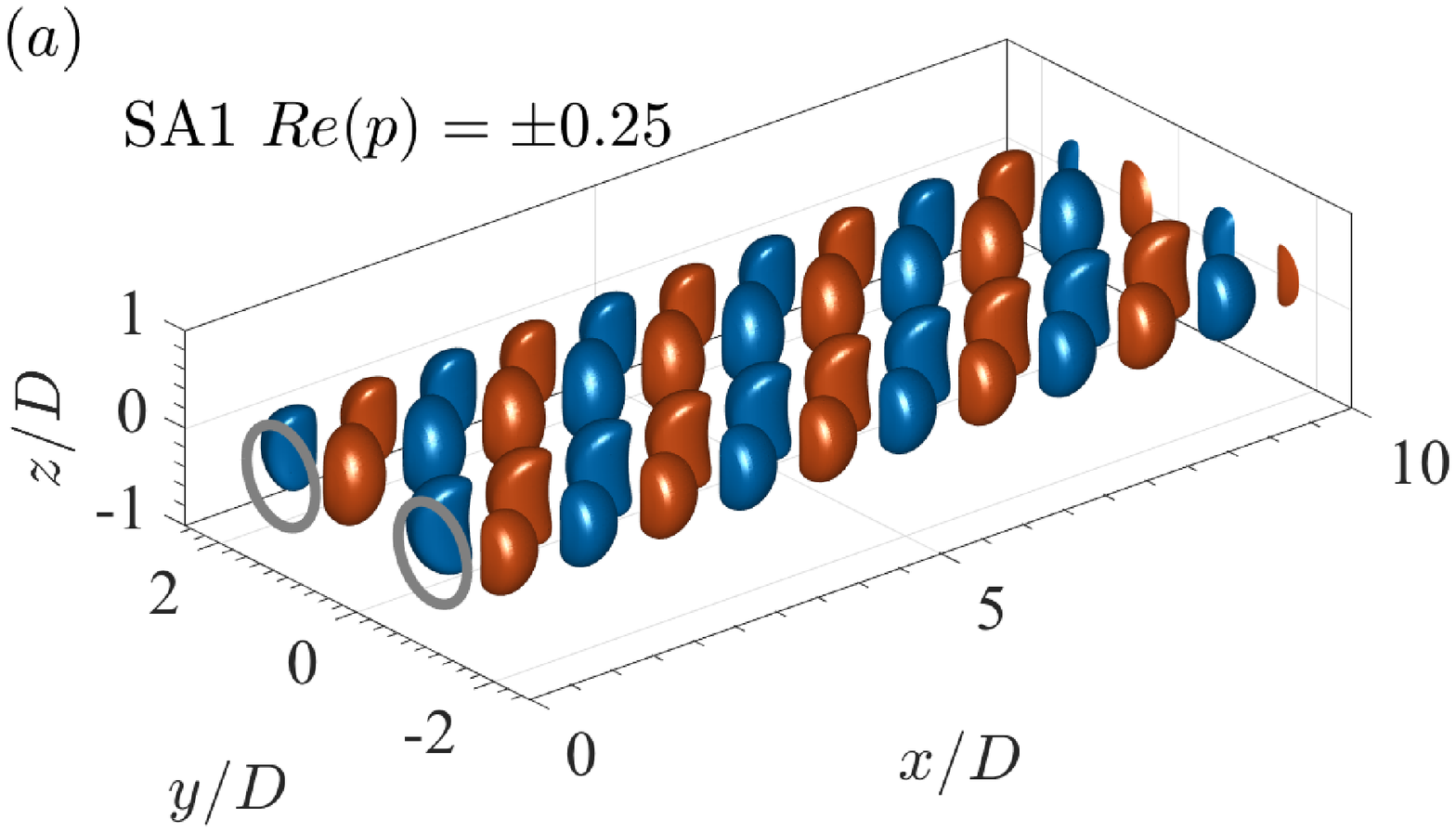} & 
\includegraphics[width=.45\textwidth]{EigVec3D_h2p2_SA1_phase} \\
\includegraphics[trim=0cm 1.5cm 0cm 2.5cm, clip=true, width=.55\textwidth]{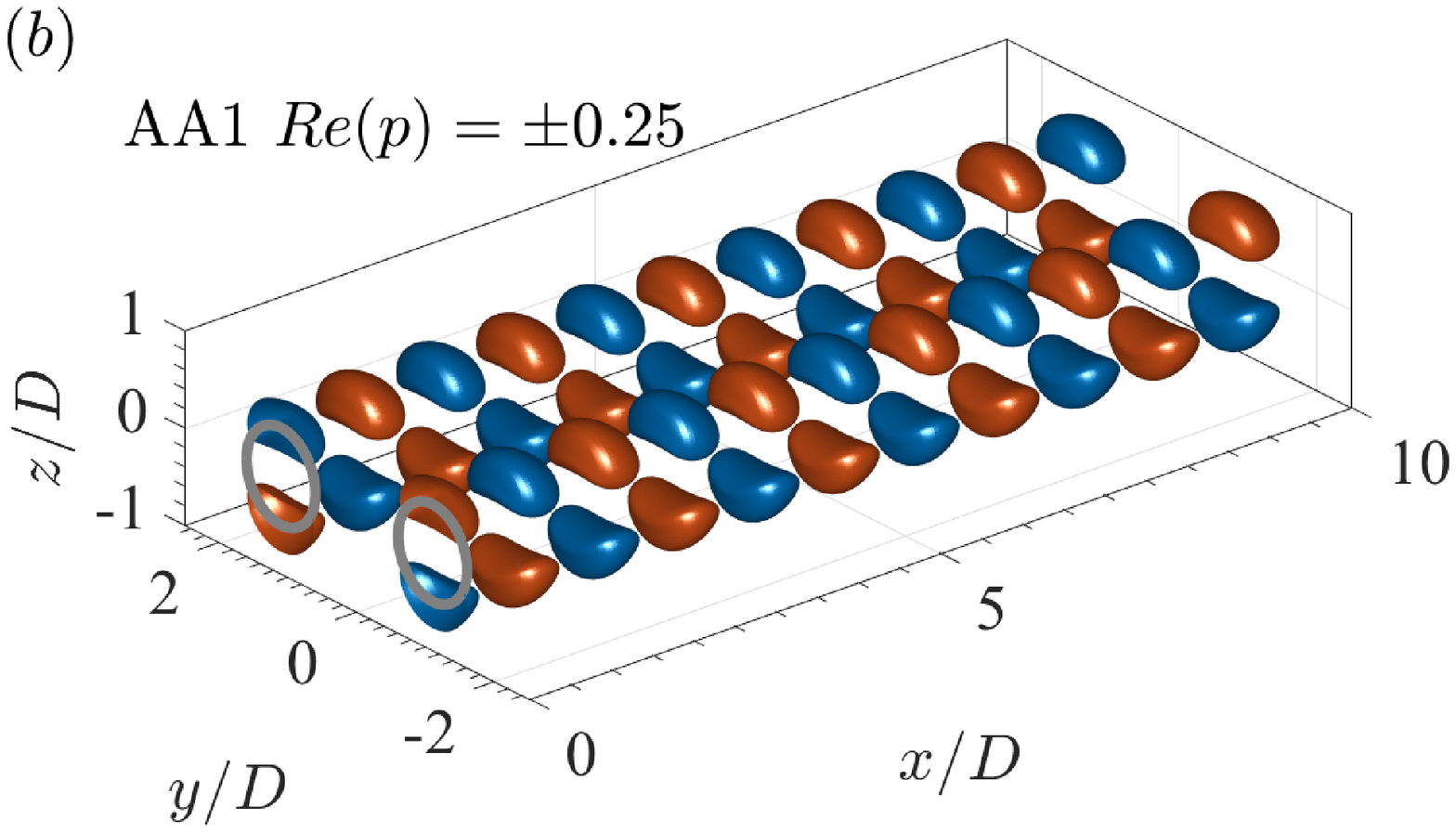} & 
\includegraphics[width=.45\textwidth]{EigVec3D_h2p2_AA1_phase} \\
\includegraphics[trim=0cm 1.5cm 0cm 2.5cm, clip=true, width=.55\textwidth]{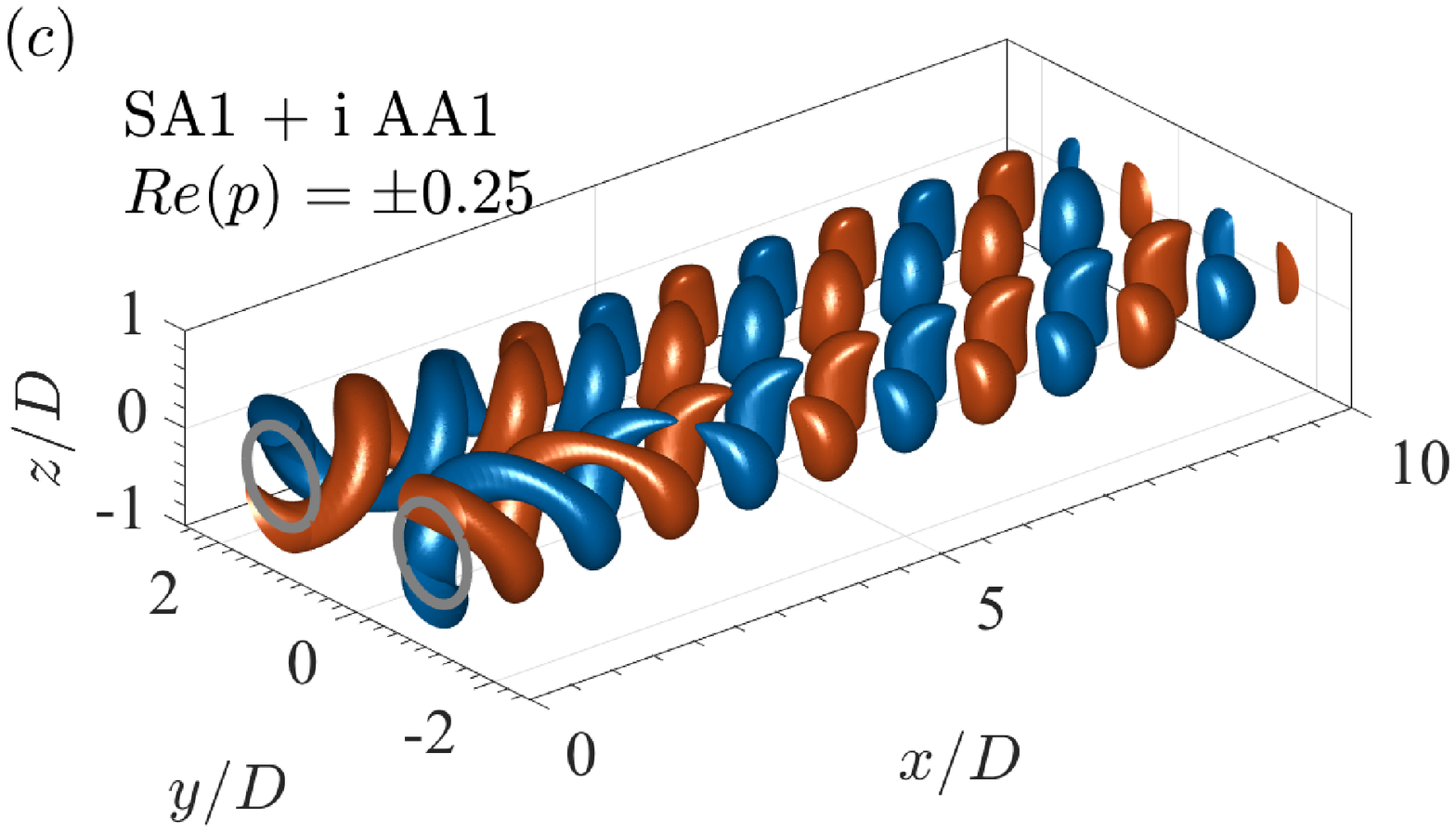} & 
\includegraphics[width=.45\textwidth]{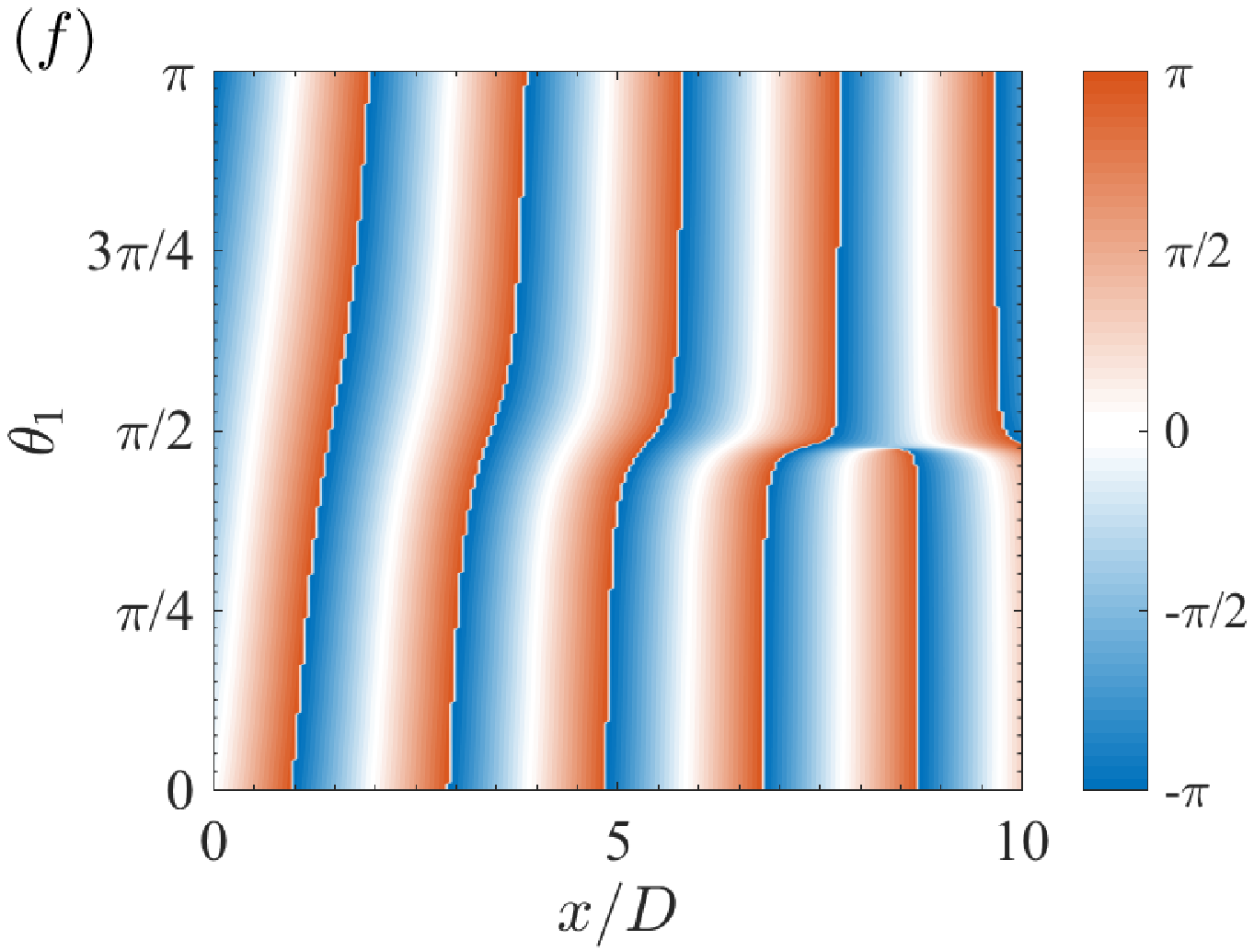} \\
\end{tabular}}
\caption{Pressure eigenfunctions corresponding to twin jets with $s/D = 2.2$ at $M_j=1.5, T_R = 1, R/\theta = 12.5$ and $St = 0.3$. The streamwise dependence is obtained from the corresponding eigenvalue, eliminating the spatial growth corresponding to SA1. Left column: iso-contours of the real pressure component. The grey circles shows the nozzle circumference. Right column: phase angle $\phi$ at a cylinder of radius 0.75$D$ centred on one jet.($a$) SA1; ($b$) AA1; ($c$) Linear combination of SA1 and AA1. }
\label{fig:EigVec_h2p2D_helical}
\end{figure}

The eigenvalue shift resulting from the coupling of the unsteady twin-jet pressure fields has two inter-connected consequences. First, some of the $m=1$ eigenmodes become dominant over the others for a given combination of parameters ($s/D, St$, but also $M_j$ and $T_R$, as shown later), suggesting that a preferred mode of flapping oscillation should emerge, as opposed to the helical motion of a single jet. From the results of figure \ref{fig:EigVal_m1}, mode SS1 (varicose lateral flapping) should be expected for jet separation above $s/D \approx 3$ and $St = 0.3$. For $s/D < 3$, mode SA1 (sinuous lateral flapping) becomes gradually dominant. 
The second consequence of the eigenvalue shift is that $m=1$ helical modes cannot be recovered as the linear combination of different $m=1$ modes. Owing to the exponential dependence on $k_i$ (see equation \ref{eqn:Results_SJ_helical}), even small differences in the spatial growth rate result in a fast dominance of the most unstable flapping eigenmode. This is illustrated in figure \ref{fig:EigVec_h2p2D_helical} for a jet separation of $s/D=2.2$. As was done for figure \ref{fig:EigVec_SJ_helical}, the streamwise amplification of the eigenfunctions is re-scaled with the growth rate of the most unstable one (SA1 in this case). As a consequence, the least amplified mode AA1 has its relative amplitude reduced along the streamwise direction. A phase mismatch also appears due to the small differences in $k_r$. The combination of the two modes is shown in figure \ref{fig:EigVec_h2p2D_helical}($c,f$): the pressure field behaves initially as two anti-symmetric counter-rotating helical modes, but soon evolves towards the sinuous lateral flapping of mode SA1.

\subsection{Preferred mode of oscillation for finite-thickness twin jets}
\label{sec:Results2}

\begin{figure}
\centerline{\begin{tabular}{cc}
\multicolumn{2}{c}{\includegraphics[width=.45\textwidth]{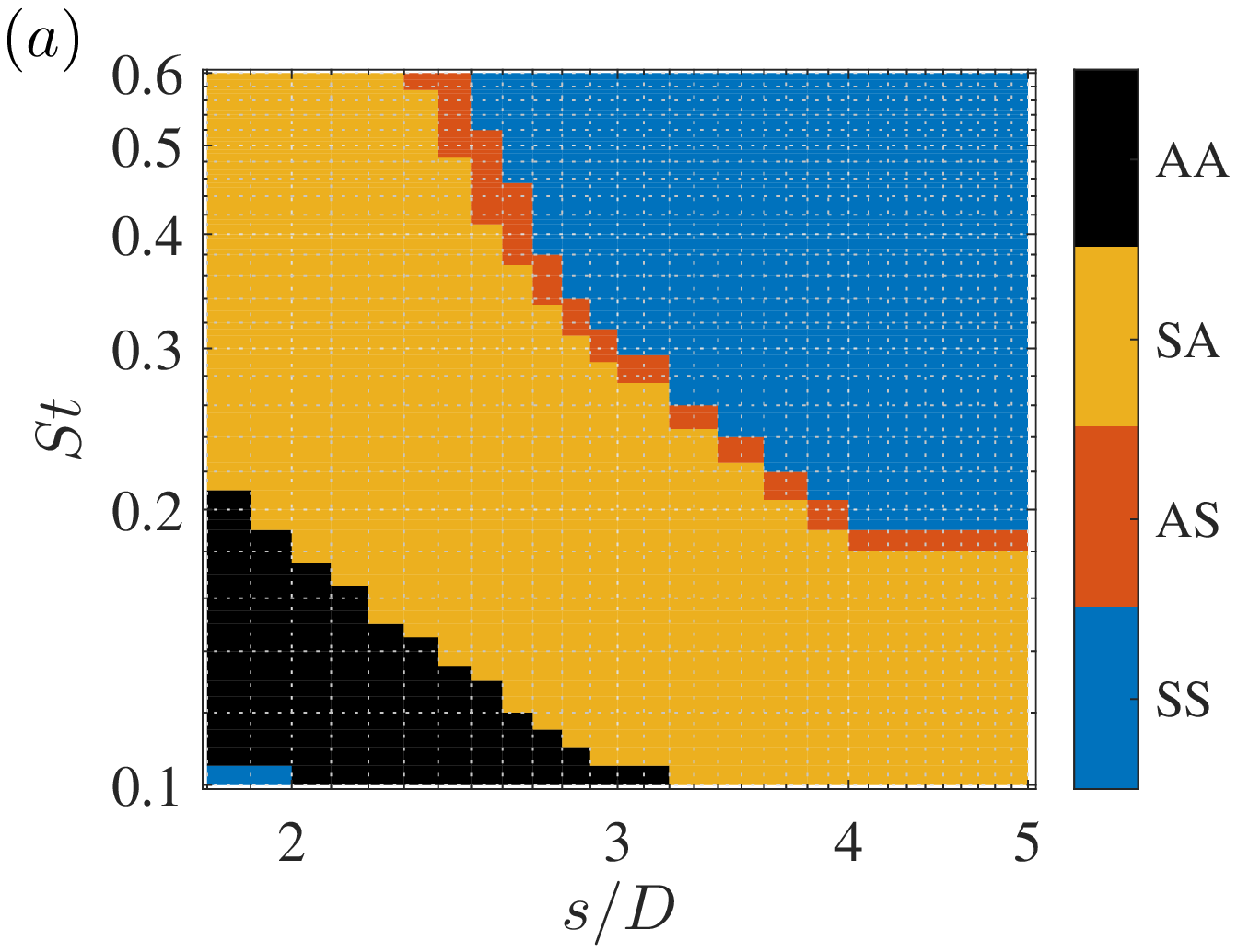}} \\
\includegraphics[width=.45\textwidth]{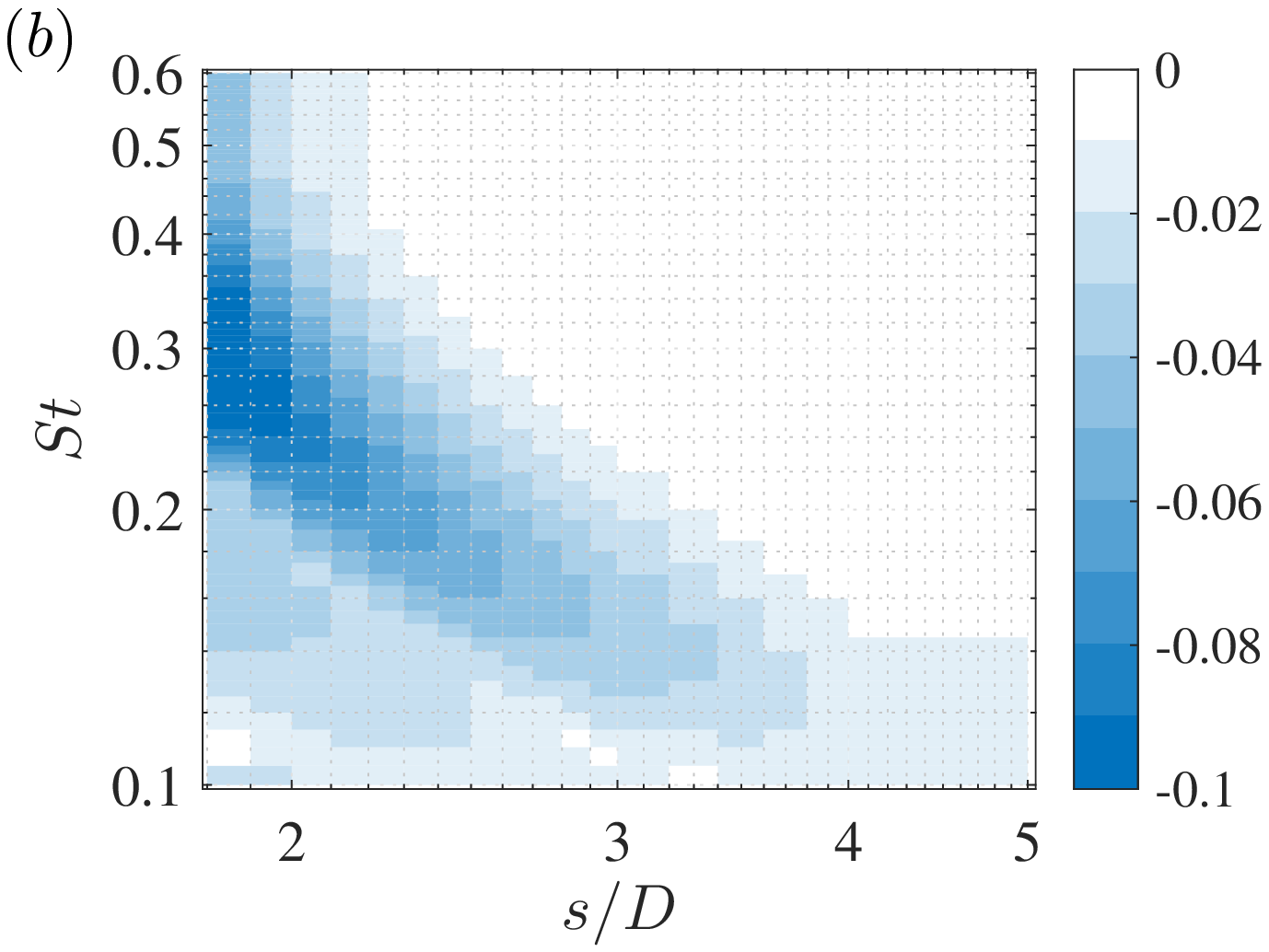} &
\includegraphics[width=.45\textwidth]{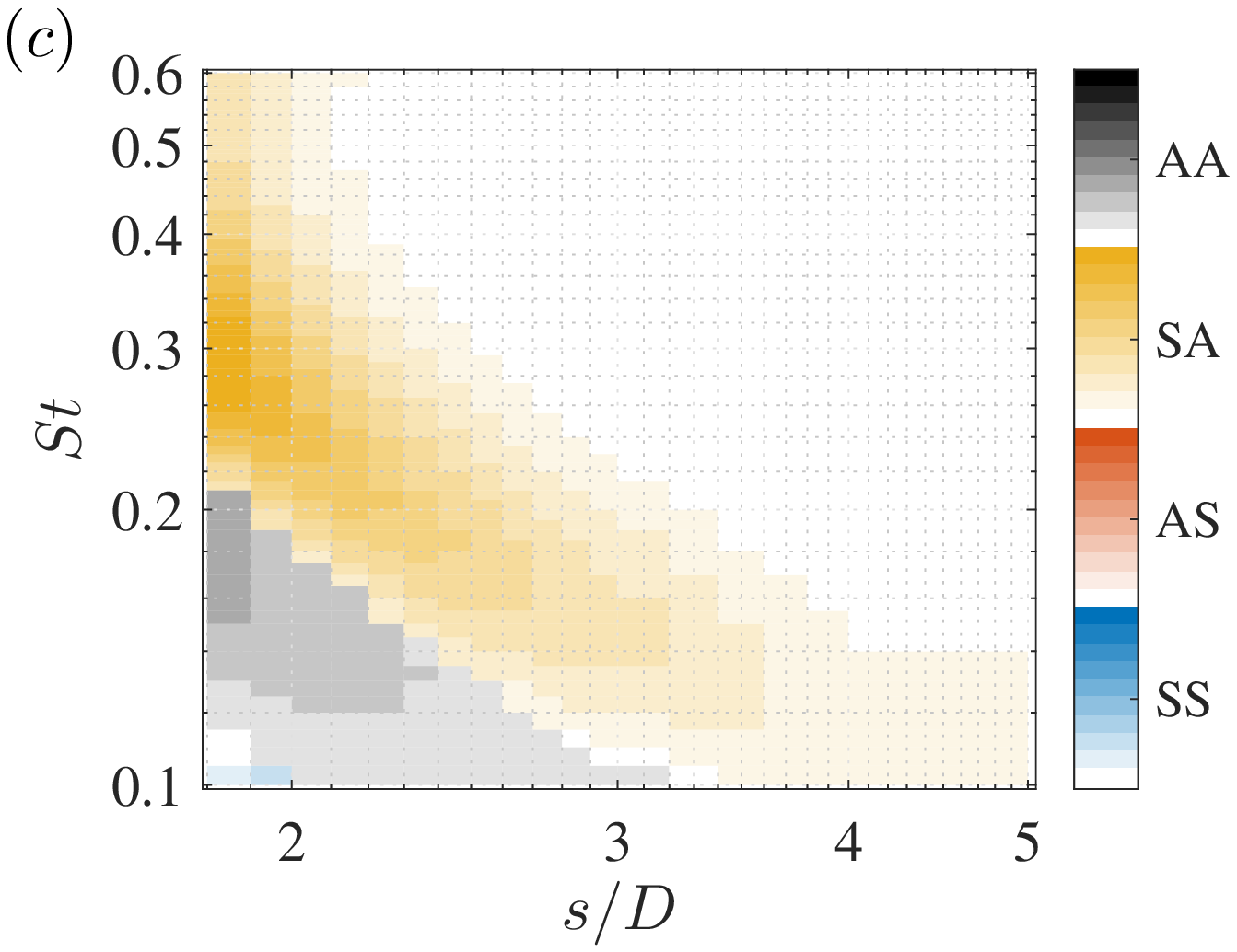} \\ 
\end{tabular}}
\caption{Preferred oscillation mode for twin jets as a function of the jet separation and Strouhal number. $M_j=1.5, T_R = 1, R/\theta = 12.5$. ($a$) Leading eigenmode. ($b$) Relative increase of the growth rate with respect to the single jet. ($c$) Same as ($b$), but colour-coded to show the leading eigenmode. The same colour-coding of figure \ref{fig:EigVal_m1} is used: Blue: SS1; Red: AS1; Yellow: SA1; Black: AA1.}
\label{fig:LST1}
\end{figure}

The results in the previous section suggest a strong dependence of the eigenvalue shift for all modes with frequency and jet separation. Figure \ref{fig:LST1} shows the results of a parametric study of the $m=1$ modes for $St$ varying between 0.1 and 0.6 and $s/D$ varying between 1.8 and 5, extending the results of figure \ref{fig:EigVal_m1} for the twin jets with $M_j=1.5$, $T_R = 1$, and $R/\theta = 12.5$.
Figure \ref{fig:LST1}($a$) shows the parametric regions in which each mode dominates, i.e. its spatial growth rate is the largest among the $m=1$ modes. Incidentally, it is noted that the growth rate of all $m=1$ is always larger than that of the $m=0$ modes, as expected for high $M_j$ jets with relatively thin shear layer \citep{MichalkePrAS1984}. The alternating dependence of the leading mode with $s/D$ described before is also obtained with $St$, and the region of preferrence of the modes forms ``stripes'' whose boundaries take approximately constant values of the product $St \times s/D$.

Figure \ref{fig:LST1}($a$) provides no information on the relative growth rates of the leading mode with respect the others, or with respect to eigenmodes corresponding to a single jet. Figure \ref{fig:LST1}($b$) shows, for the leading eigenmode at each $(St,s/D)$, the relative change in the growth rate resulting from the jet interaction, quantified as 

\begin{equation}\label{eqn:ev_shift}
\Delta k_i (St,s/D) = \dfrac{k_i(St,s/D) - k_i(St,s/D \to \infty)}{|k_i (St,s/D \to \infty)|}.
\end{equation}

Note that the dispersion relation for twin jets $k(St,s/D)$ recovers that of a single jet in a smooth manner as $s/D \to \infty$. Each colour level in figure \ref{fig:LST1}($b$) corresponds to a relative increase of 1\% of $\Delta k_i$. The maximum colour level shown is 10\% but $\Delta k_i$ reaches larger values. As anticipated, the destabilisation of the leading eigenmode is inversely proportional to jet separation. 
For a fixed jet separation, the envelope of $|\Delta k_i|$ over all modes is also inversely proportional to $St$, similarly to the dependence on $s/D$ shown in figure \ref{fig:LST1}(d).
Consequently, as the product $St \times s/D$ increases, $\Delta k_i$ approaches zero. This happens in this case for regions dominated by the modes SS1 and AS1. Figure \ref{fig:LST1}($c$) shows again $\Delta k_i$, but colour coded to highlight the preferred oscillation mode. For the jet conditions analysed herein ($M_j=1.5, T_R = 1, R/\theta = 12.5$), these results predict the dominance of mode SA1 leading to lateral sinuous oscillations for most jet separations and Strouhal numbers. The white regions in figure \ref{fig:LST1}($c$) correspond to conditions in which the jet interactions are weak, and hence do not lead to a substantial eigenvalue shift; the oscillation modes converge towards their behaviour for single jets for which there is no preferred mode. In this case, the jet may follow either helical of flapping dynamics, which may be selected by other characteristics of the turbulent flow. 

The results presented so far consider finite-thickness jets and are obtained using the linear stability analysis formulation described in section \ref{sec:Method1}. While those calculations are performed using a personal computer, the parametric study leading to figure \ref{fig:LST1} requires the numerical solution of over 10 000 individual matrix eigenvalue problems, which becomes impractical for more comprehensive studies including the effect of the jet Mach number and the temperature ratio. In the following section, such a study is performed using the (zero-thickness) twin-jet shear layer model presented in section \ref{sec:Method2}.

\begin{figure}
\centerline{\begin{tabular}{cc}
\includegraphics[width = .45\textwidth]{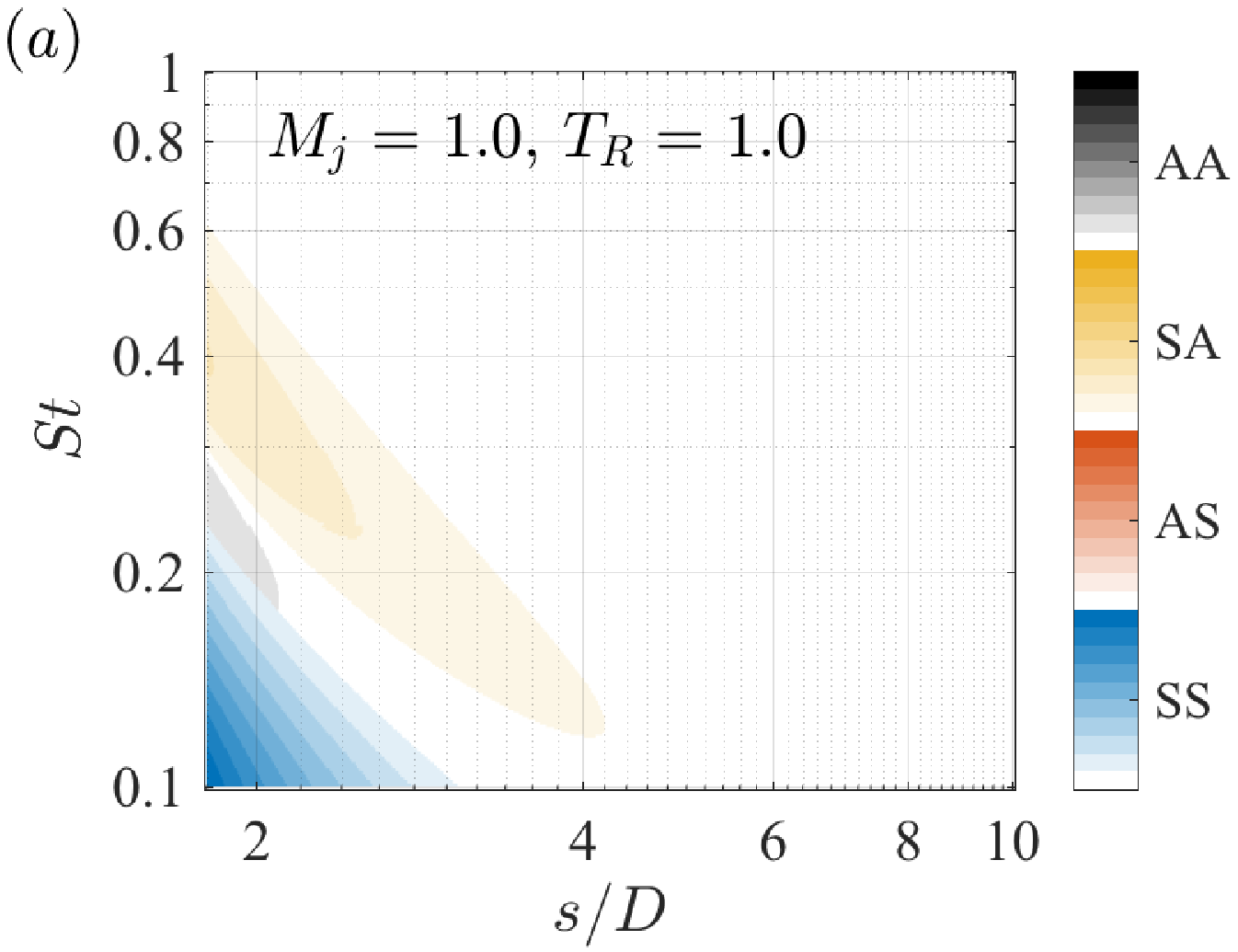} & 
\includegraphics[width = .45\textwidth]{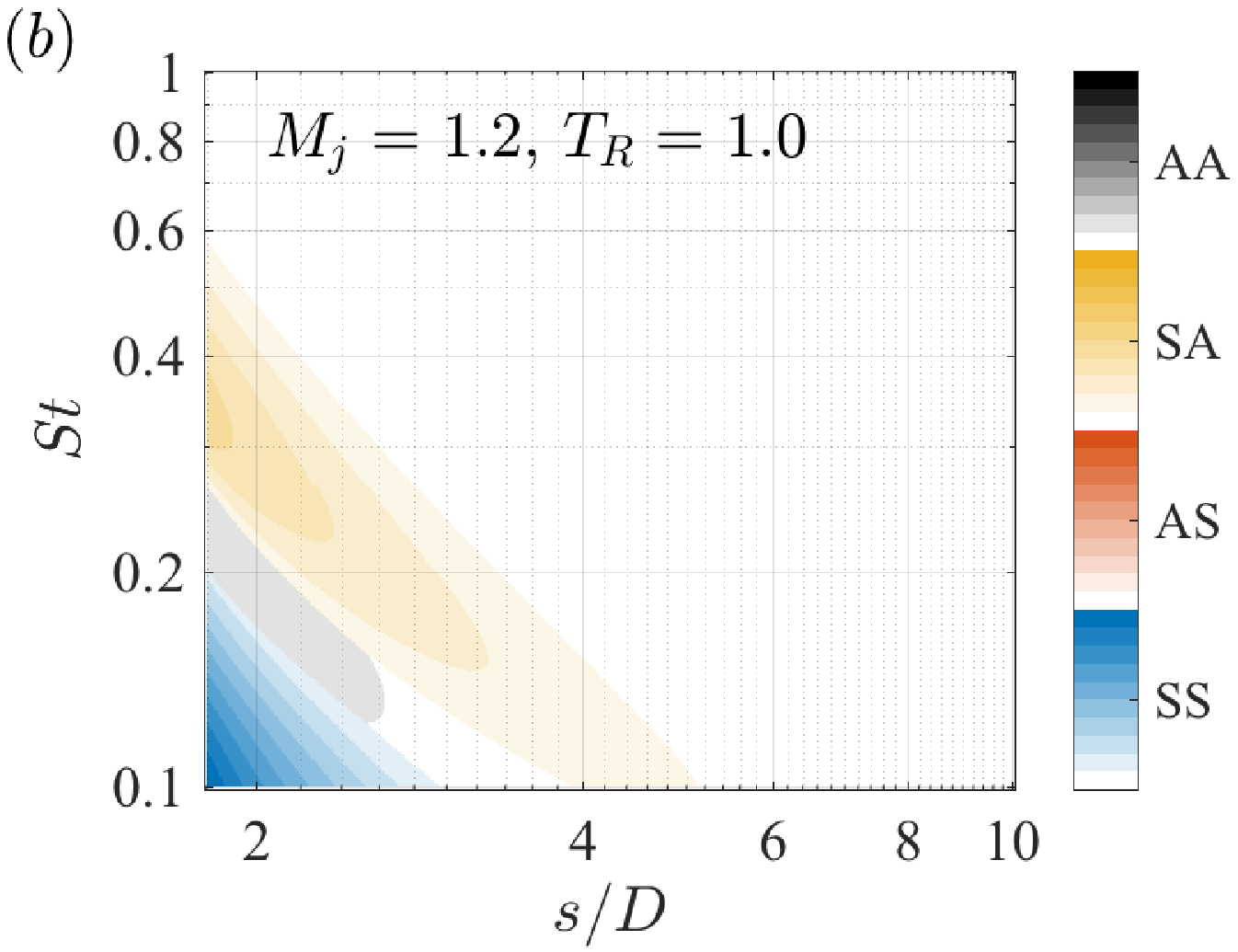} \\
\includegraphics[width = .45\textwidth]{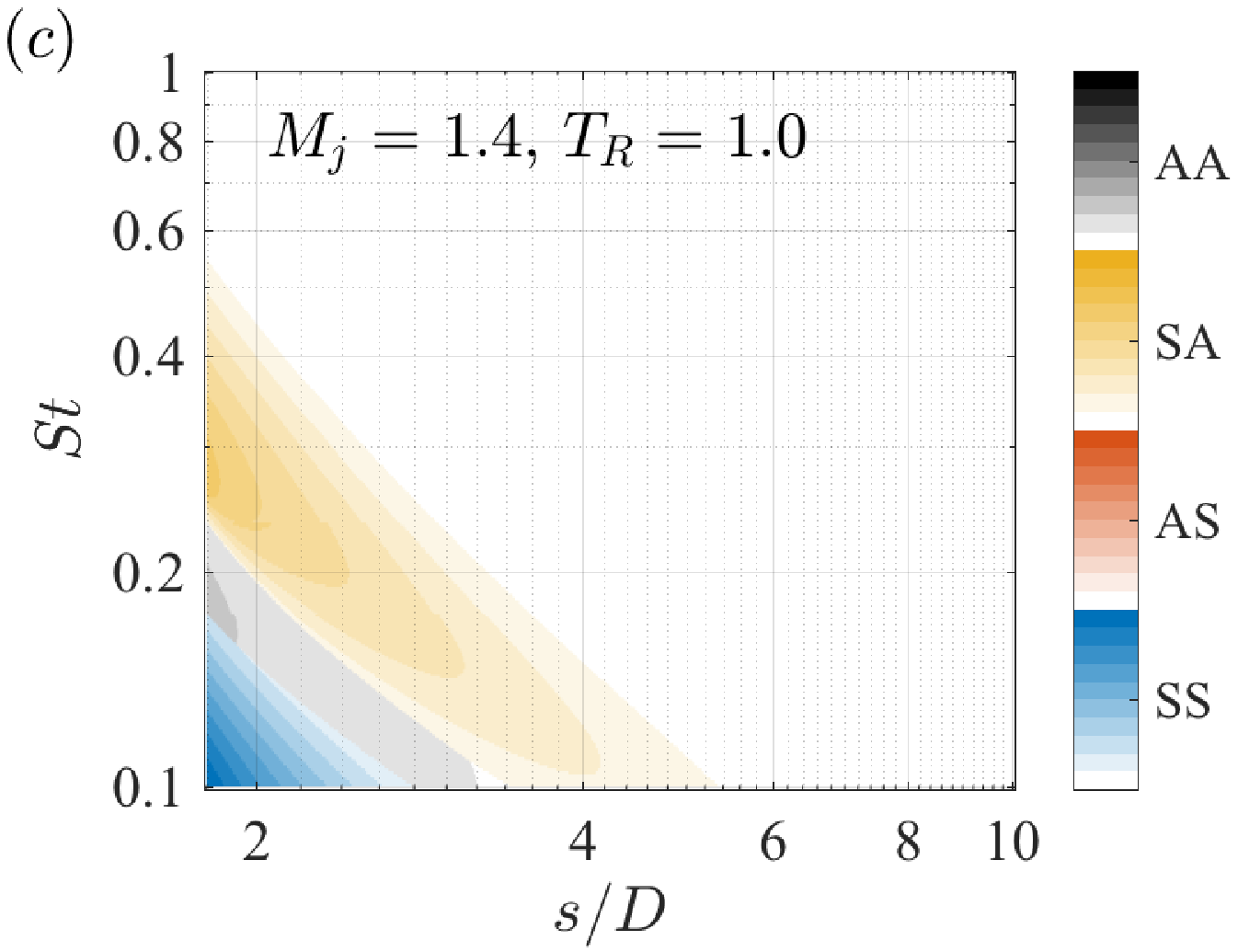} & 
\includegraphics[width = .45\textwidth]{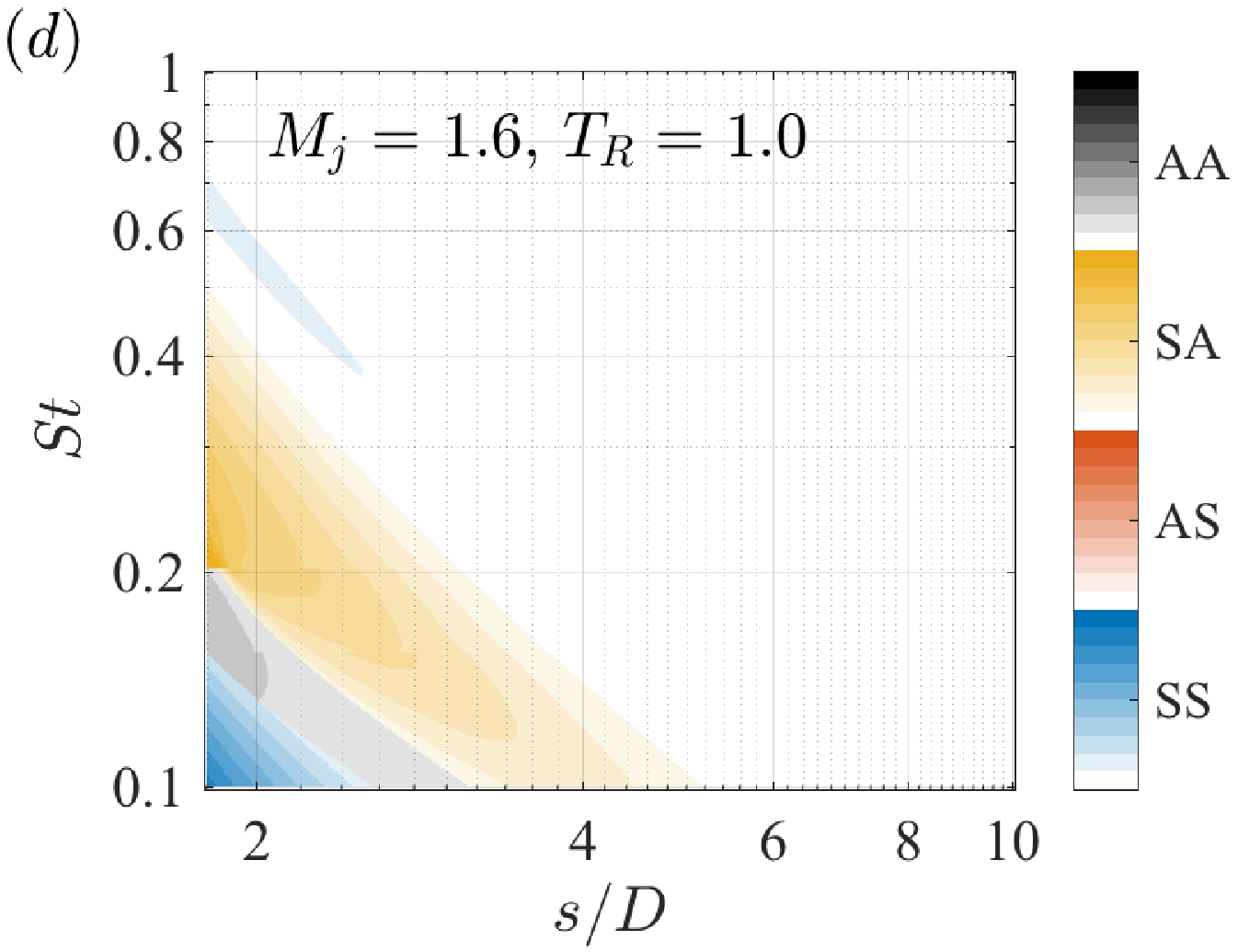} \\
\includegraphics[width = .45\textwidth]{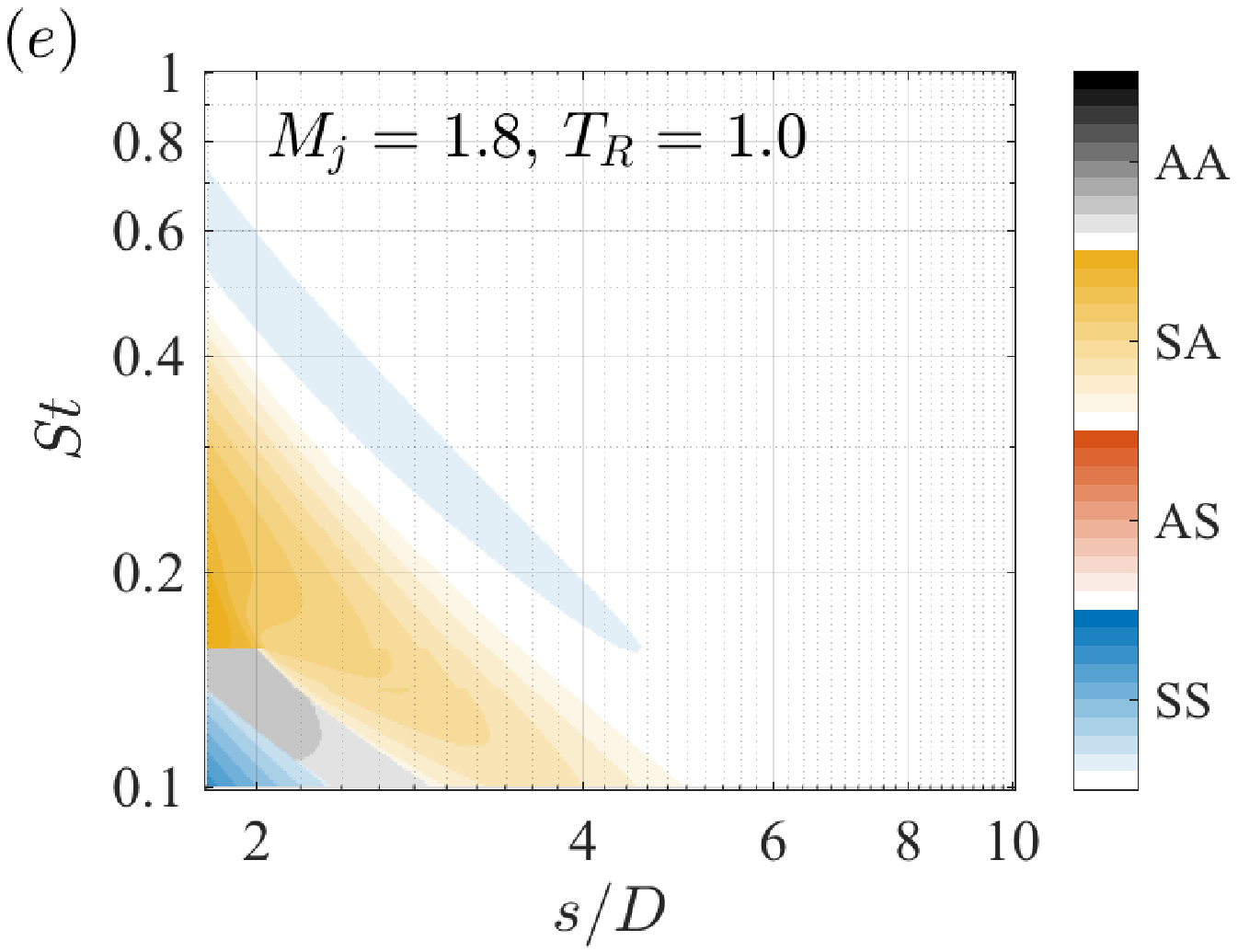} & 
\includegraphics[width = .45\textwidth]{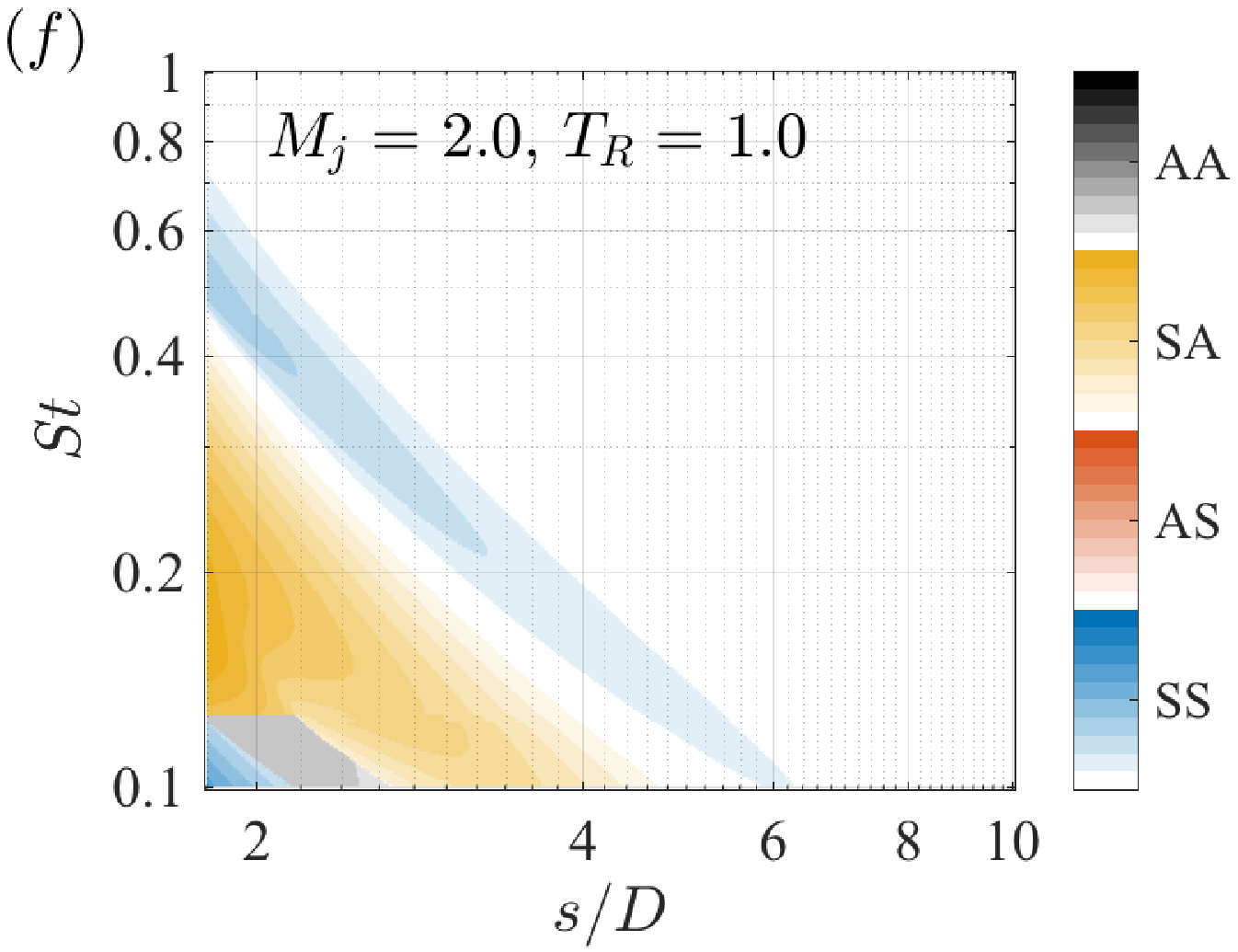} 
\end{tabular}}
\caption{Preferred oscillation mode for twin jets as a function of the jet separation and Strouhal number. Vortex-sheet model, $T_R = 1$ and varying $M_j$. Contours as in figure \ref{fig:LST1}($c$). Blue: SS1; Red: AS1; Yellow: SA1; Black: AA1.}
\label{fig:TwinVS_M_imd_dkir}
\end{figure}

\begin{figure}
\centerline{\begin{tabular}{cc}
\includegraphics[width = .45\textwidth]{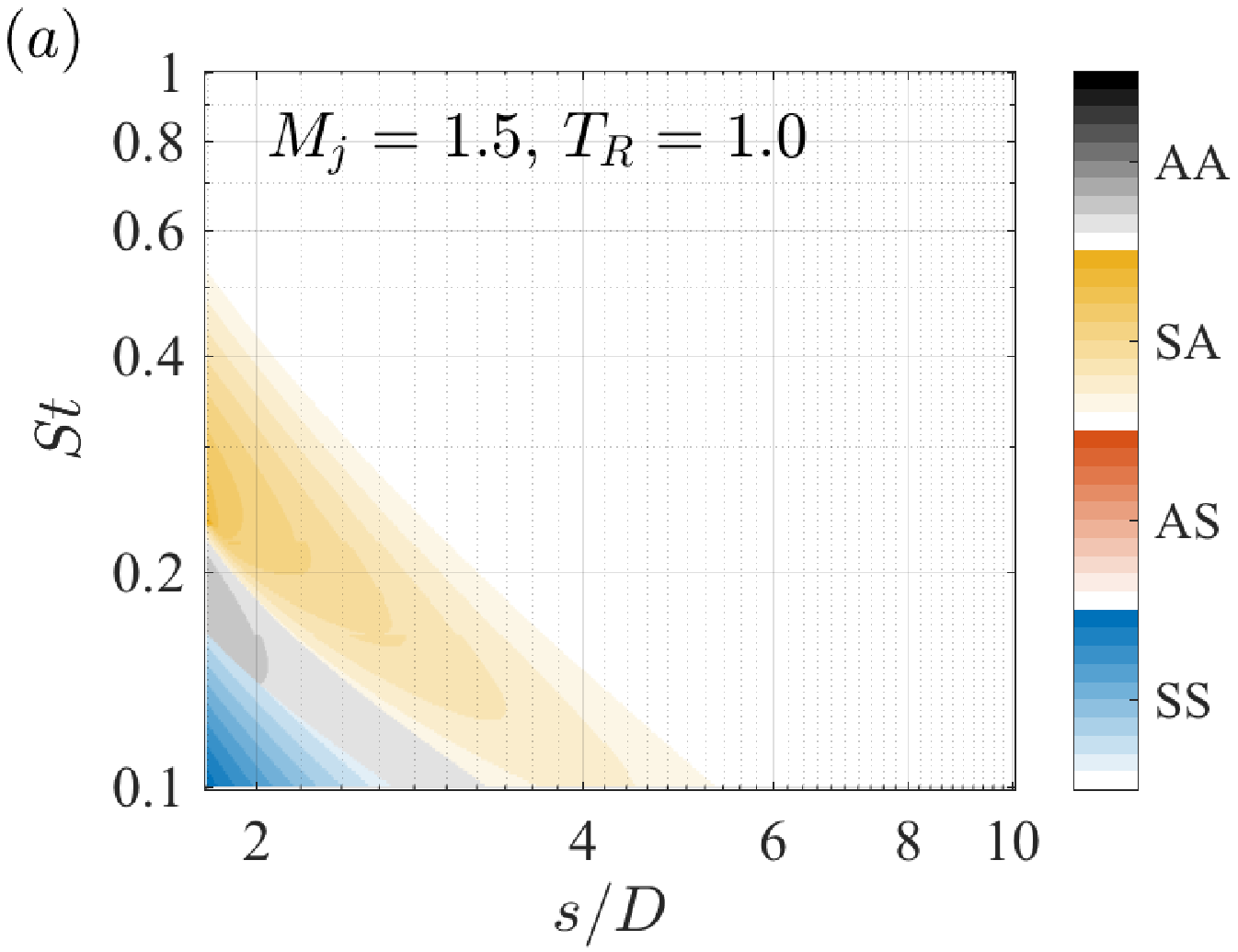} & 
\includegraphics[width = .45\textwidth]{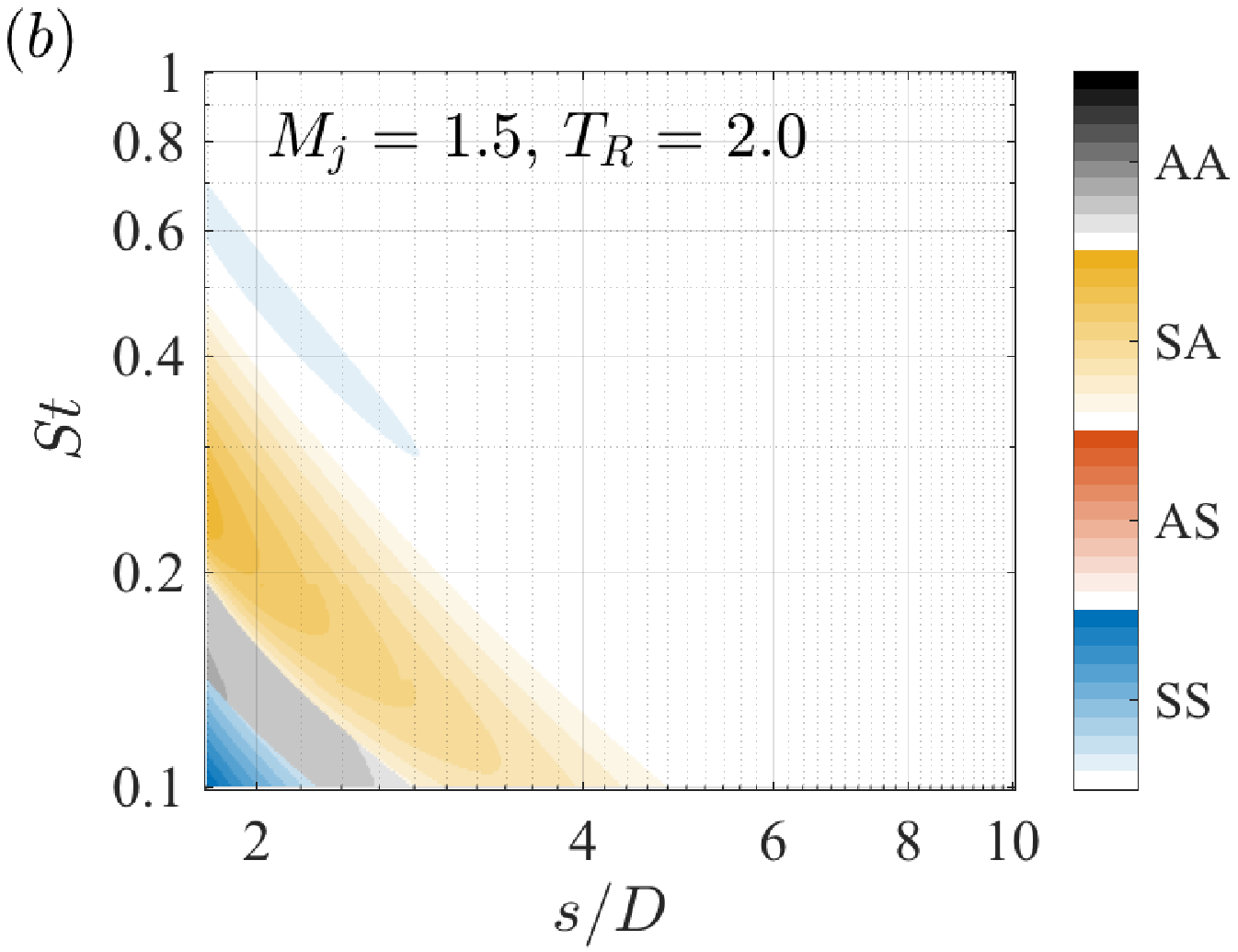} \\
\includegraphics[width = .45\textwidth]{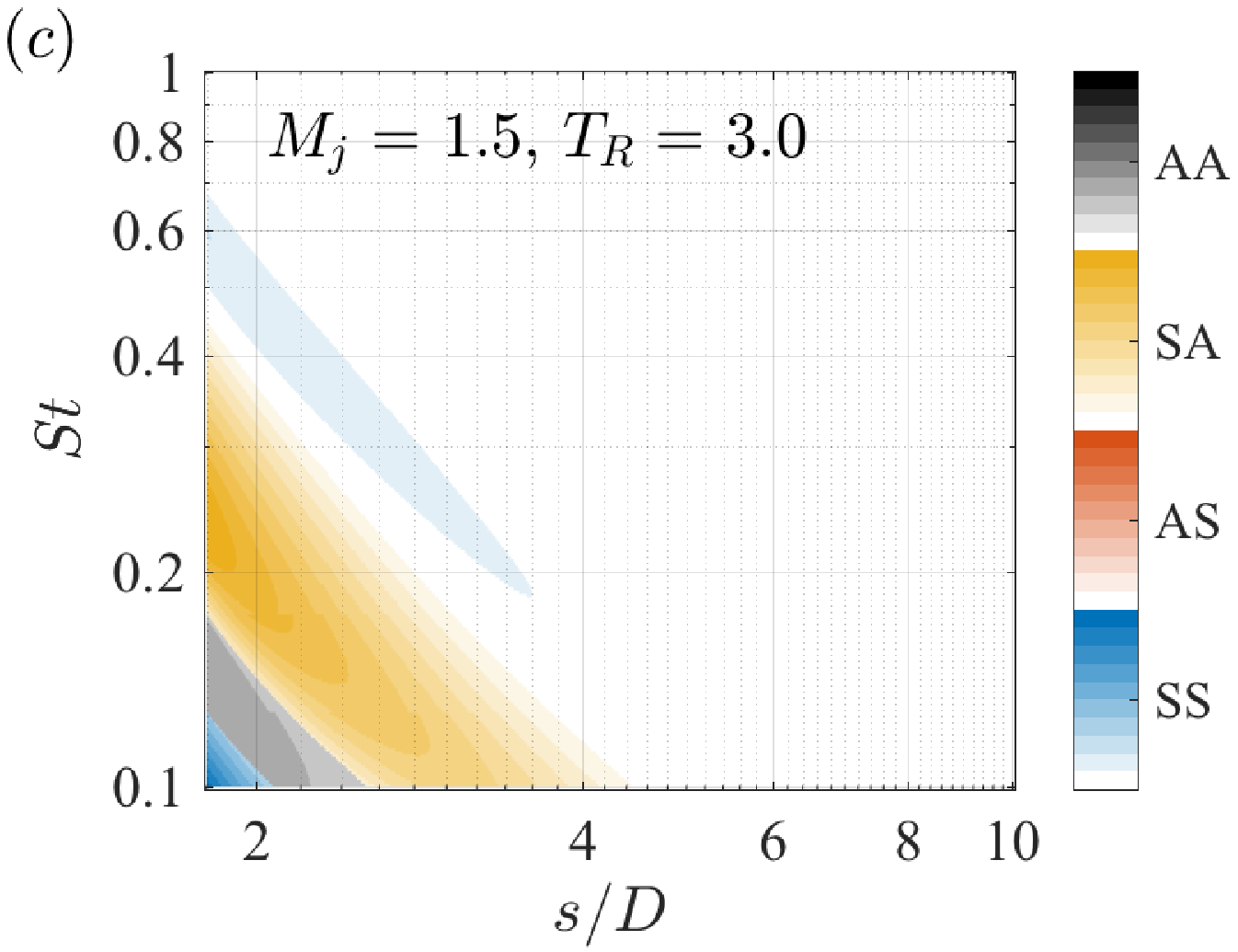} & 
\includegraphics[width = .45\textwidth]{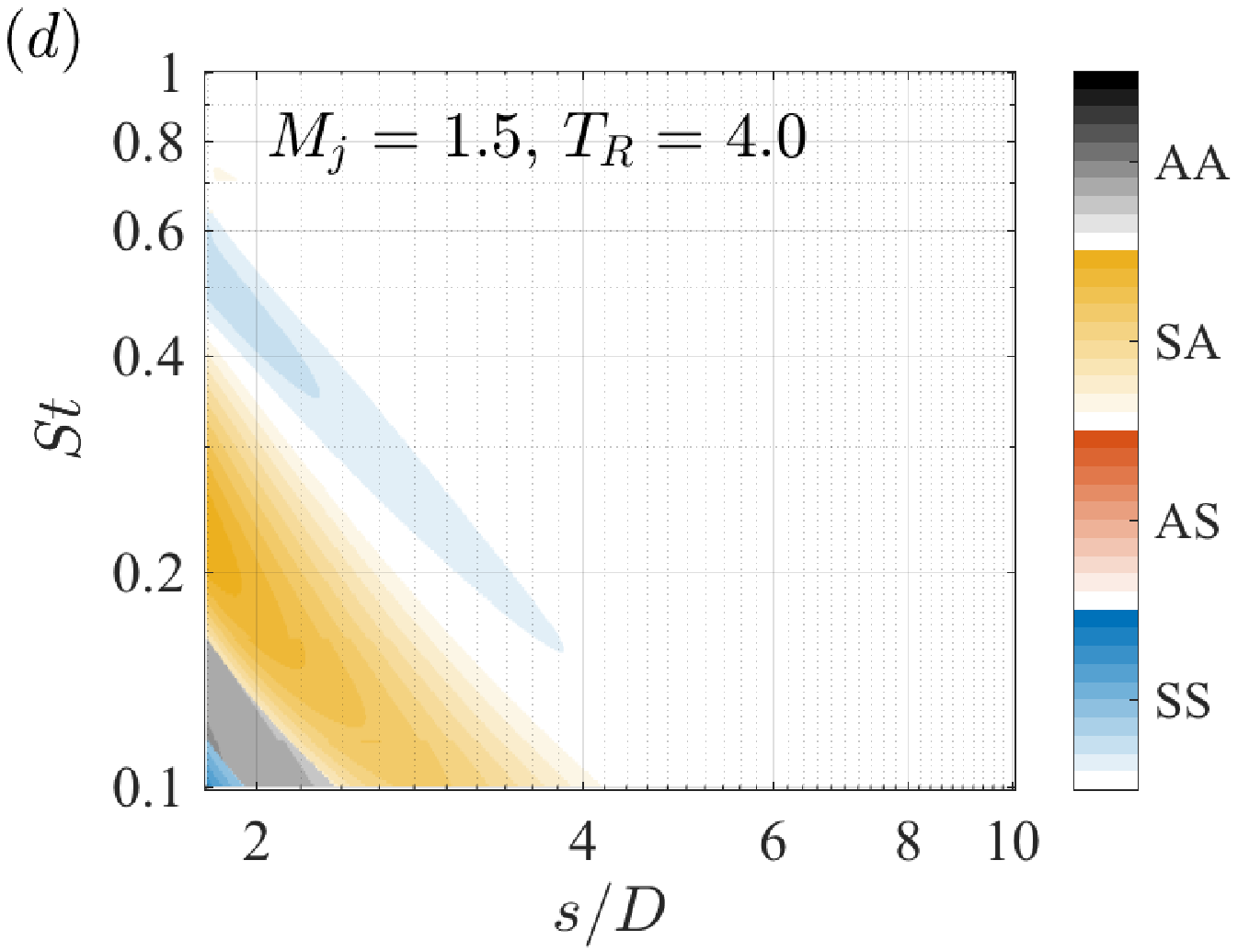} 
\end{tabular}}
\caption{Preferred oscillation mode for twin jets as a function of the jet separation and Strouhal number. Vortex-sheet model, $M_j = 1.5$ and varying $T_R$. Contours as in figure \ref{fig:LST1}($c$). Blue: SS1; Red: AS1; Yellow: SA1; Black: AA1.}
\label{fig:TwinVS_T_imd_dkir}
\end{figure}

\subsection{Dependence of the preferred mode of oscillation on Mach number and temperature ratio}
\label{sec:Results3}

This section presents a parametric study of the effect of the jet Mach number $M_j$ and temperature ratio $T_R$ on the preferred oscillation mode, using the vortex-sheet model for twin round jets. The description of the eigenmode families, the eigenvalue shift and their qualitative dependence on jet separation and Strouhal number given for finite-thickness jets holds for the vortex-sheet model. For each $(M_j,T_R)$, the evolution of the four $m=1$ eigenmodes with jet separation and Strouhal number is monitored in a manner analogous to that described in the previous section. The reduced computational cost of the vortex-sheet model computations allows for a greater region of interest and to significantly refine the discretisation of the $(St,s/D)$ space. The analysis for one $(M_j,T_R)$ case involves the calculation of nearly 1.5 million eigenvalues, but takes less than an hour on a modern personal computer.

Figure \ref{fig:TwinVS_M_imd_dkir} shows the leading oscillation mode for twin jets with $T_R = 1$ and $M_j$ between 1 and 2. The overall picture reproduces that already described for the finite-thickness $M_j = 1.5$ case:
 
(i) the regions in which each particular mode dominates in the $(St,s/D)$ space form alternating stripes; 

(ii) the maximum eigenvalue shift, computed following (\ref{eqn:ev_shift}), occurs for the smallest jet separation within each strip; 

(iii) as the jet separation is reduced, the Strouhal number for the maximum shift of the growth rate is increased. 

For $M_j = 1$ (figure \ref{fig:TwinVS_M_imd_dkir}($a$)) and $s/D \le 3$, mode SS1 dominates for $St \le 0.2$. For larger $St$, modes SA1 and AA1 become dominant but their eigenvalue shift is comparatively weak. Increasing the jet Mach number at constant temperature ratio has two effects. First, the boundaries of the mode stripes are gradually displaced towards lower $St$ and $s/D$ values. Second, the shift of the growth rates is increased. The latter implies that the interaction between the fluctuation fields of the two jets intensifies with $M_j$. The combination of the two effects leads to the gradual appearance of a new stripe of mode SS1 at higher $St$ while the corresponding stripe at the lower $St$ region reduces. Simultaneously, the stripe of mode SA1 is displaced to lower Strouhal numbers while its growth rate shift is increased. For $M_j$ above 1.5, mode SA1 (sinous lateral oscillations) has the largest growth rate shift.

For $M_j$ above 1.6 (figures \ref{fig:TwinVS_M_imd_dkir}($d-f$)), the results for mode SA1 present an anomalous behaviour in the limit of lower separation and Strouhal number, that is clearly visible in the figures. Instead of following the general trend, the boundary of the stripe forms a horizontal segment (constant $St$) for $s/D$ below a threshold value increases with $M_j$. A preliminary study revealed that this anomaly is due to the appearance of a saddle point in the evolution of mode SA1, which pinches with mode SA2 at a certain $(St,s/D)$ combination. This saddle point is not associated with an absolute instability, as it involves two downstream-propagating K-H eigenmodes. A deep study regarding this issue is currently underway but is beyond the scope of this paper. 

Figure \ref{fig:TwinVS_T_imd_dkir} shows the parametric study for twin jets with fixed jet Mach number $M_j = 1.5$ and temperature ratio increasing from $T_R = 1$ to 4. The effect of increasing the jet temperature follows the same trends as increasing the jet Mach number: the mode stripes move towards lower $St$ and $s/D$, reducing the region dominated by mode SS1 (dominant for $T_R = 1$) and increasing the relative importance of mode SA1 with increasing temperature. Overall, the impact of the temperature ratio on the eigenmode shift is weaker than that of the jet Mach number, but still mode SA1 attains growth rates nearly 10\% larger that those for the equivalent single jet.


\section{Conclusions}
\label{sec:Conclusions}

Linear stability theory is often used to model the large-scale flow structures in the turbulent mixing region and near pressure field of high-speed jets. For perfectly-expanded single round jets, these models predict the dominance of $m=1$ helical modes for relatively low frequencies and thin shear layers, in agreement with empirical data. 

For twin-jet configurations, the interaction between their flutuation fields results in coupling. The symmetries present in the geometry lead to a separation of the modes into four families, corresponding to the even/odd behaviour of the pressure field about the symmetry plane and the plane containing the jet axes. The division into families is inherent to the twin-jet geometry and not just an artifact of the problem formulation: while it is exploited here to simplify the analyses, a formulation in which the symmetries/anti-symmetries are not imposed recovers the same families \citep{Rodriguez:CRM2018}. The fluctuation field associated with $m = 1$ modes for twin jets does not correspond to helical oscillations, but to flapping oscillations of the jet. 

The coupling between the jets is quantified by monitoring the shift of the complex wavenumber $k$ (eigenvalue of the spatial linear stability problem) with respect to that of the single jet, $\Delta k$. The absolute value of $\Delta k$ is different for each oscillation mode, but for all of them is inversely proportional to both the jet separation $s/D$ and the Strouhal number $St$. However, the real and imaginary parts of $\Delta k$ do not present a monotonous dependence on $St$ or $s/D$. This leads to regions in the $(St,s/D)$ space in which different modes of oscillation dominate over others. A parametric study is presented in which the jet Mach number $M_j$ is varied between 1 and 2, and the jet temperature ratio $T_R$ is varied between 1 and 4. Increasing independently both $M_j$ and $T_R$ is found to augment the jet coupling and modify the $(St,s/D)$ map of the preferred oscillation mode. Present results predict that only two oscillation modes are likely to be observed when jet coupling is relevant, namely in-phase (varicose) or counter-phase (sinuous) flapping oscillations on the plane containing the nozzles (modes SS1 and SA1, respectively)

It should be noted that while flapping modes are the natural mode of oscillation of twin jets, a superposition of them may still generate some helical motions in regions close to the nozzle. However, owing to the differences between the growth rates of each mode and their exponential amplification in the first few diameters, it can be expected that flapping motions will dominate the flow as it develops downstream. Hence, the eventual helical oscillations in twin-jet systems would be a spatially transient feature of the flow. 

The results presented in this work consider perfectly expanded jets. The preference of certain oscillation modes over others is predicted based on the larger growth rate of the corresponding Kelvin-Helmholtz instability eigenmode.
It may be expected that these results would be applicable also to shock-containing jets, at least to some extent. The preference of shock-containing twin jets to present flapping motions at conditions in which a single jet presents helical motions has been shown by different authors \citep{Seiner:AIAAJ88,Wleizen:AIAAJ89,Kuo:AIAAJ17,Knast:AIAAJ18,Bell:EF18,Bell:JFM21}.
However, one should be cautious when extending the present conclusions to shock-containing jets. 
As these jets are subjected to the resonance loop associated with screech \citep{Powell:53,Edgington:IJAero19,nogueira2022absolute}, the actual dominant oscillation is a function of the symmetry associated with the most globally unstable resonance mode. As shown in \citet{Tam:JSV1982,Edgington:JFM21,Nogueira:JFM21,nogueira2022closure,edgington2022unifying}, the closure mechanism of screech is strongly dependent on the wavenumber of the waves involved in resonance, and less dependent on the growth rate of the Kelvin-Helmholtz instability. However, as the wavenumber of the Kelvin-Helmholtz eigenmode is one of the driving components of this phenomenon, the present results remain relevant: they may be used as input for screech models in twin jets and help explain the preferred flapping mode in shock-containing jets as well.


\section*{Acknowledgements}
D.R. is funded by the Government of the Community of Madrid within the multi-annual agreement with Universidad Polit\'ecnica de Madrid through the Program of Excellence in Faculty (V-PRICIT line 3) and the Program of Impulse of Young Researchers (V-PRICIT lines 1 and 3, Grant No. APOYO-JOVENES-WYOWRI-135-DZBLJU).
M. N. S. is supported through an Australian Government Research Training Program Scholarship.
P.A.S.N. and D.M.E.M. were supported by the Australian Research Council through the Discovery Project scheme (Grant No. DP190102220).

\bibliographystyle{jfm}
\bibliography{MyBibliography}


\end{document}